\documentclass[12pt]{article}
\usepackage{epsfig}
\newcommand{\z}{&&\hspace*{-1cm}}

\newcommand{\as}{ \overline \alpha_{s}}

\newcommand{\bea}{\begin{eqnarray}}
\newcommand{\eea}{\end{eqnarray}}

\begin{document}

\setcounter{page}{0}
\thispagestyle{empty}


~\vspace{2cm}

\begin{center}
{\Large {\bf Analytic continuation
of the Mellin moments \\
of deep inelastic structure functions }} \\[0pt]
\vspace{1.5cm} {\large \ A.V. Kotikov } \\[0pt]
\vspace{0.5cm} {\em Bogoliubov Laboratory of Theoretical Physics \\[0pt]
Joint Institute for Nuclear Research\\[0pt]
141980 Dubna, Russia }\\[0pt]

\vspace{0.5cm} and \\[0pt]
\vspace{0.5cm} {\large \ V. N. Velizhanin }\\[0pt]
\vspace{0.5cm} {\em Theoretical Physics Department\\[0pt]
Petersburg Nuclear Physics Institute\\[0pt]
Orlova Roscha, Gatchina\\[0pt]
188300, St. Petersburg, Russia }\\[0pt]
\end{center}

\vspace{1.5cm}\noindent

We derive the analytic continuation
of the Mellin moments of deep inelastic structure functions
at the next-to-next-to-leading order accuracy.

\newpage
\pagestyle{plain}

\section{Introduction}

Deep inelastic (lepton-hadron) scattering (DIS) is
one of the best studied
reactions now.
It provides unique information about the structure of the hadrons and
tests one of the most important predictions of perturbative QCD, the
scale evolution of the structure functions (SF) \cite{GroWi}-\cite{AnNonPol}.

The increasing accuracy of the DIS experiments demands more
accurate theoretical predictions. Very recently the calculations of the 3-loop
corrections to anomalous dimensions (AD) of Wilson operators
have been performed in \cite{MVVns,MVVsi} that leads to complete theoretical
information needed to analyze inclusive
DIS reactions at the next-to-next-to-leading
order (NNLO) accuracy.

The results have been presented in the Bjorken $x$-space for the corresponding
splitting functions and also in the momentum space (i.e. in $n$-space)
for the anomalous
dimensions themselves. Although the $x$-space results have been done in
complete form, the results for the anomalous dimensions have been presented
in the form of nested sums, which are correct only for even or odd
values of moments and
cannot be used, for example, to determine
directly
the exact values of various sum rules, which correspond in the nonsinglet case
to the Mellin moments of structure functions at $n=1$.

Of course, the sum rules can be extracted directly by integration of
the $x$-space results
for the splitting functions (see, for example, \cite{RoSa}-\cite{BroKaMa}).
However, it is
convenient to have a proper
representations for the SF Mellin moments, where the sum rules values can be
obtained automatically at $n=1$.

Despite the sum rules, the correct $n$-space
representations are important
also to reconstruct the structure functions and/or parton distributions
from their corresponding moments. In general, for the back Mellin
transformation someone should know
the Mellin moments for the complex $n$ values.

To study the structure functions at intermediate $x$ values, sometimes there
are
important only the integer values of the Mellin moments. It is the case,
for example, for programs to fit DIS
experimental data, which are based on the Bernstein and Jacobi polynomials
(see \cite{YndPL}
and \cite{PaSo,Krivo}, respectively).
The programs are based on the exact solution of the DGLAP evolution equations
\cite{DGLAP}
in the momentum space and on the reconstruction of the DIS structure functions at
the end of evolution with help of the orthogonal polynomials
(see, for example, \cite{Escob},\cite{SaYnd} and \cite{Krivo},
\cite{Vovk}-\cite{Kataev}
for the Bernstein and Jacobi polynomials, respectively).

This procedure is simpler to compare with the numerical solution of the
DGLAP equations in $x$-space, which is used usually in global fits of
experimental data (see, for example, \cite{fits} and references therein).
The simplicity leads to possibility to use only partial information
about the DIS coefficient functions
and/or anomalous dimensions.
For example, the first NNLO analysis of $F_2$ and $F_3$ structure functions
have been obtained in \cite{PKK} and \cite{Kataev}
just after the first several even and odd moments of the
nonsinglet anomalous dimensions have been calculated in \cite{LRV}.

To have the analytic continuation it is important also to study a similarity
between the DGLAP \cite{DGLAP} and BFKL \cite{BFKL} equations in the framework
of ${\mathcal N}=4$ Supersymmetric
Yang-Mills (SYM) theory (see \cite{KoLi03}). The analytic structure of the
BFKL kernel \cite{KoLi00} in this model gives the possibility to predict the
eigenvalues of AD matrix at the first three orders of perturbation theory
(see Refs. \cite{Li}, \cite{KoLi03} and \cite{KLOV}, respectively, and
discussions therein). At the first two orders the predictions have been checked
by direct calculations in \cite{KLV}.

Following to the studies \cite{LoYn}-\cite{KoPRD94}, it is possible to show that
at small $x$ values the $Q^2$-evolution of DIS structure functions is
described (see
\cite{Kot}) by the behavior of their moments with the ``number''
$n=1+\delta$ in the case of Regge-type asymptotics of SF, i.e., for example,
 $F_2(x) \stackrel{x \to 0}{\sim} x^{-\delta}$. In this case the continuation
of the SF moments to real
$n$ numbers is already needed.

The analytical continuation has been already obtained in \cite{KaKo,Ko94}
(see also some similar studies in \cite{GRV}),
for a quite simple set of the nested sums $S_{-a,b,c, ...}(n) $
(hereafter $a,b, c, ... $ are integer and positive)
\footnote{Below we will consider the positive and negative integer values
for the first three symbols of the nested sums. The values of other symbols
are marked by $ ... $ and
taken be always integer and positive.}:
\begin{eqnarray}
S_{-a,b,c, ...}(n)~=~ \sum^n_{k=1} \frac{(-1)^k}{k^a} \, S_{b,c, ...}(k),~~
 \label{I0}
\end{eqnarray}
where
\begin{eqnarray}
 S_{b,c, ...}(n)~=~ \sum^n_{k=1} \frac{1}{k^b} \, S_{c, ...}(k),~~
S_{d}(n)~=~ \sum^n_{k=1} \frac{1}{k^d}.
 \label{I0ad}
\end{eqnarray}

Such nested sums contribute to the NLO corrections to the anomalous
dimensions and coefficient functions (see
\cite{AnNonPol,Bardeen,KaKo,KaKoTMF}).
At the NNLO level, the QCD anomalous
dimensions and coefficient functions \cite{MVVns,MVVsi,MoVe1,MVV02}
contain more complicated nested sums
$S_{\pm a,\pm b,\pm c, ...}(n) $
\footnote{We would like to note that at ${\mathcal N}=4$ SYM model the
corresponding NNLO
anomalous dimensions can be expressed in the form (\ref{I0}) of the nested
sums (see \cite{KLOV}).}
and their continuation is the main subject of the study.

We note that the set of the nested sums is not fully independent:
there are a lot of relations between the nested sums (see, for example,
\cite{Blumlein} and references therein) and, as the basic ones, we can use
for the NNLO anomalous dimensions
only ones in Eq.~(\ref{I0}) and the additional sum $S_{a,-b,c, ...}(n) $.
Thus, really it is necessary to apply the results of \cite{KaKo,Ko94} and
to study only the sums  $S_{a,-b,c, ...}(n) $.

Note, however,  that the form of the
NNLO
anomalous dimensions are quite cumbersome and the new representation
containing  the  nested sums (\ref{I0}) and $S_{a,-b,c, ...}(n) $
will be long, too.
So, it is better to keep the original Moch-Vermaseren-Vogt (MVV)
representations and to give the analytic continuation for each nested
sum of the  MVV results.

Thus, the purpose of this study is the extension of the procedure of the
analytic continuation given in \cite{KaKo,Ko94} for the
more complicated nested sums $S_{\pm a,\pm b,\pm c, ...}(n) $, that
needs to $n$-space representations for anomalous dimensions  and coefficient
functions, which should be
correct for arbitrary $n$ values. Moreover, after continuation the $n$-space
representations should have
the form which is very close to the original one in
\cite{MVVns,MVVsi,MoVe1,MVV02}.
As a by product of the study we present
the results for $n=1$ which are related with QCD sum rules.

The structure of the paper is a following.
Section 2 contains the general information about the properties of DIS
structure functions and about method to extract the results for coefficient
functions and anomalous dimensions.
In section 3 we reproduce the basic steps of the analytic continuation
\cite{KaKo,Ko94}.
Section 4 contains the results of the analytic continuation for all needed
nested sums $S_{\pm a,\pm b,\pm c, ...}(n) $.
In sections 5 and 6 there are some examples of the application of the analytic
continuated results. Conclusion contains a summary of our results.
Appendices A and B contain the basic steps of the procedure of the analytic
continuation of the nested sums
to the integer and real/complex arguments, respectively.

\section{Basic formulae}

The optical theorem relates the DIS structure functions to the forward
scattering amplitude
of photon-nucleon scattering, $T_{\mu\nu}$, which has a time-ordered
the product of two local electromagnetic currents, $j_{\mu}(x)$ and
$j_{\nu}(z)$.
After Fourier transformation to momentum space, the standard
perturbative theory can be applied. The operator product expansion
allows to expand this current product around the light-cone
$(x-z)^2 \sim 0$ into a series of local composite operators
$O_{\mu_1, ... , \mu_n}$ of leading
twist and spin $n$.
The anomalous dimensions on matrix elements
$<h|O_{\mu_1, ... , \mu_n}|h> = p_{\mu_1, ... , \mu_n} A_n(p^2/\mu^2)$
\footnote{ Here $p$ is hadron moment, $\mu^2$ is the renormalization scale,
which is equal to the factorization scale in our study (see below Eqs. (\ref{I01}),
(\ref{I1})
and (\ref{I2})) and $p^{\mu_1,...,\mu_n}$ is traceless product (see its
definition and properties, for example, in \cite{CheKaTka}).}
of these operators govern the
scale evolution of the structure functions, while the coefficient
functions
multiplying these matrix elements are calculable
in perturbative QCD.

Thus, for the scalar structure functions $T_{i}$ $(i=2,L,3)$ of the forward
scattering amplitude $T_{\mu\nu}$ we have
\begin{eqnarray}
T_{i} (Q^2) &=&
\sum_{n=0}^{\infty}
\frac{1}{x^n} T_{i,n} (Q^2), \nonumber \\
T_{i,n} (Q^2) &=&
\sum_{k=NS,q,g} C^k_{i,n} (Q^2/\mu^2,\alpha_s) A^k_{n}(\mu^2),
~~~~(i=2,3,L).
\label{I01}
\end{eqnarray}

The Wilson operators
$A^k_{n}(\mu^2)$ denote the spin-averaged hadronic matrix elements
and $C_i$ are the coefficient functions and the sum extends over the flavor
non-singlet and singlet quark and gluon contributions.

In this way the Mellin moments of DIS structure functions can naturally
be written in the parameters of the operator product expansion
(here and below our $F_3$ structure function is equal to standard
$xF_3$ function)
\begin{eqnarray}
\frac{1+(-1)^n}{2} \, \int^1_0 dx x^{n-2} F_i(x,Q^2) ~=~
\sum_{k=NS,q,g} C^k_{i,n} (Q^2/\mu^2,\alpha_s) A^k_{n}(\mu^2)
~~~~(i=2,3,L)
\label{I1}
\end{eqnarray}
for $F^{e(\mu)p}_{2,L}(x,Q^2)$
in the electron(muon)-proton scattering
and for $F^{\nu p + \overline \nu p }_2(x,Q^2)$
and \\
$F^{\nu p - \overline \nu p }_3(x,Q^2)$
in  the (anti-)neutrino-proton scattering and
\begin{eqnarray}
\frac{1-(-1)^n}{2} \, \int^1_0 dx x^{n-2}
F_i(x,Q^2) =
C^{NS}_{i,n} (Q^2/\mu^2,\alpha_s) A^{NS}_{n}(\mu^2)
~~~~(i=2,3)
 \label{I2}
\end{eqnarray}
for $F^{\nu p - \overline \nu p }_2(x,Q^2)$ and
$F^{\nu p + \overline \nu p }_3(x,Q^2)$
in the (anti-)neutrino-proton scattering.

The difference in Eqs. (\ref{I1}) and (\ref{I2}) comes from the
relations $F^{e(\mu)p}_{2,L}(-x)=F^{e(\mu)p}_{2,L}(x)$,
$F^{\overline \nu p }_2(-x)=-F^{\nu p }_2(x)$ and
$F^{\overline \nu p }_3(-x)=F^{\nu p }_3(x)$ based on the
charge symmetry (see \cite{Buras} and references therein).

From Eqs. (\ref{I1}) and (\ref{I2}) one can see that only even
and odd moments of the structure functions
can be calculated from
the odd (even) coefficients of the expansion of the
corresponding functions $T_{2}$, $T_{L}$ and $xT_3$.

Thus, when we
used $x$-space splitting functions coming in the forward
scattering amplitude for full $x$-range,
we neglect possible terms
$\varphi(x) \pm \varphi(-x)$. This trivial analytic continuation in
$x$-space leads at the lowest order of perturbation theory to similar
trivial  analytic continuation in
$n$-space: we apply our results obtained at even or odd values to all
integer one and after trivial extension to all real and/or complex
$n$ values (see subsection 3.1).
The nontrivial case
comes at $n$-space starting at NLO approximation
when nonplanar configurations start to contribute. The configurations
lead to nonalternative nested sums which should be accurate
continued to integer, real and/or complex results starting from even or
odd ones.

Of course, after integration of the corresponding splitting-functions
we obtain automatically this analytic continuation
(see the review \cite{Altarelli} and discussion therein). However, it is
useful to have
a procedure which allows to obtain directly the
$n$-space results, completely extended to integer, real and/or
complex results.

The coefficient functions and the anomalous dimensions are
independent of a
particular
hadronic matrix element, so that it is standard to calculate partonic
structure functions with external quarks and gluons in
perturbation theory. In practice, this procedure reduces to the
task of calculating the $n^{th}$ moment of all 4-point
diagrams that contribute to $T_{\mu\nu}$ at a given order
in perturbation theory (see \cite{KaKo,KaKoTMF}).

To show this, it is better to use the Chetyrkin et al. ``gluing'' method
\cite{gluing}\footnote{In practice, the  calculation of the $n^{th}$ moment is
more convenient by an extension of ``projectors'' method
\cite{projector}  (see \cite{KaKo,KaKoTMF}).
The extension to three-loop calculations has been done in
\cite{MoVe1,VeMoch}.}.
 The method allows to extract the contribution to
coefficient function from a diagram by gluing its gluon legs by the
additional propagator having the specific form:
$q^{\mu_1,...,\mu_n}/q^{2\alpha}$, where $q$ is gluon momentum
and $\alpha$ is a special parameter.

\begin{figure}[t]
\begin{center}
\epsfig{figure=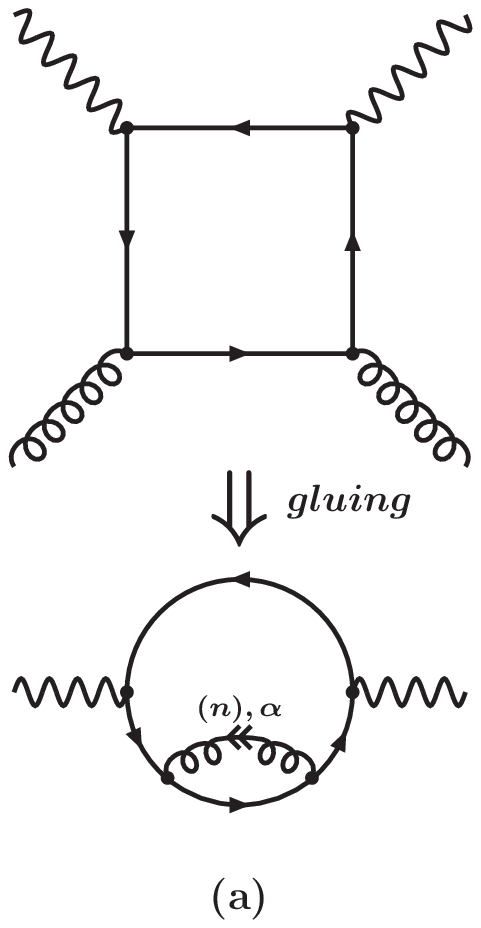,height=10cm}~~~~~~~~~~~~~~~\epsfig{figure=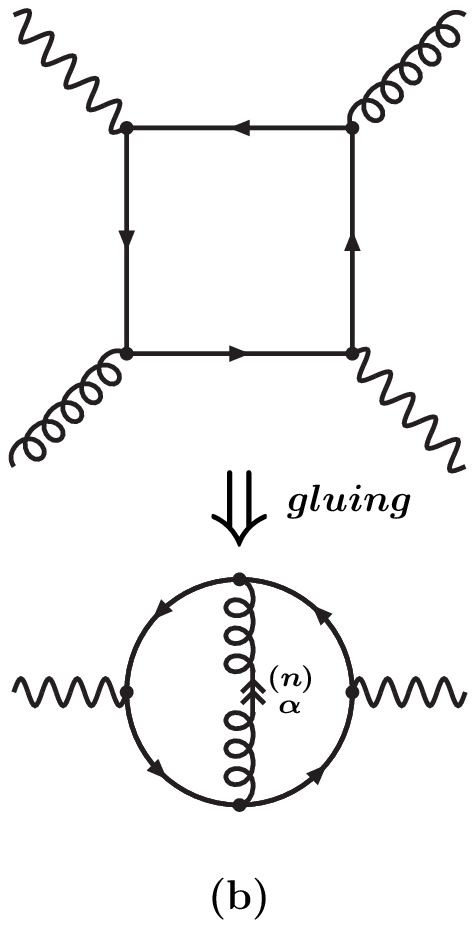,height=10cm}
\end{center}
\vspace*{-7mm}
\caption{ The diagrams contributing to $T_{\mu\nu}$ for a gluon target.
They should be multiplied by a factor of 2 because of the opposite direction
of the fermion loop. The diagram (a)
should be also doubled because of crossing symmetry.}
\label{fig1}
\end{figure}

As an example, we consider the diagrams on Fig.~\ref{fig1}, which contribute
to gluon parts of the coefficient functions and the anomalous dimensions.

The diagram (b) is
less complicated nonplanar diagram\footnote{
The diagram takes the nonplanar form if we put its legs like in the
diagram (a).},
which contributes to coefficient
functions. After gluing, the diagram obtains the new line containing
$q^{\mu_1,...,\mu_n}$.

The scalar Feynman integral
having same topology
has the following form ($k_1-k_2 \equiv q$), when $\alpha=1$,
\begin{eqnarray}
\int d^Dk_1 d^Dk_2 \frac{(k_1-k_2)^{\mu_1} ... (k_1-k_2)^{\mu_n}}{
k_1^2 k_2^2 (p-k_1)^2 (p-k_2)^2 (k_1-k_2)^2} ~=~ \hat N \,
I_2(n) \frac{p^{\mu_1, ..., \mu_n}}{{(p^2)}^{5-D}}  \, ,
\label{FI1}
\end{eqnarray}
where $\hat N$ is a normalization factor and $D=4-2\epsilon$ is a
space dimension.

When $D=4$, we have
\begin{eqnarray}
I_2(n) ~=~ \delta^0_n \cdot 6\zeta(3) -
\frac{1+(-1)^n}{n(n+1)} \, (1-\delta^0_n) \, 4 S_{-2}(n),
\label{FI2}
\end{eqnarray}
where $\delta^0_n$ is the Kroneker symbol.

The example demonstrates the fact that the sum $S_{-1}(n)$ does not
contribute to coefficient functions. The results for anomalous dimensions
support the
property. The absence of $S_{-1}(n)$ and also $\zeta(2)$
in results for coefficient functions and anomalous dimensions is sometimes
very important result. For example, it helps very much in reconstruction
of the eigenvalues of the NLO and NNLO
anomalous dimensions of ${\mathcal N}=4$ Supersymmetric Yang-Mills theory (see \cite{KoLi03} and \cite{KLOV},
respectively). Moreover, the
property can be a cross-check of the various
calculations.

Note that $I_2(n)=0$ for odd $n$ values. It is easy to see from the symmetry
property $k_1 \leftrightarrow k_2$ of the Feynman integral (\ref{FI1})
and up-down symmetry of gluing diagram on Fig.~\ref{fig1}.

The zero values of the scalar coefficients $T_{i,n}(Q^2)$ $(i=2,L,3)$
is the usual property of
forward scattering amplitude $T_{\mu\nu}$.

As an example, we consider the diagrams on Fig.~\ref{fig1}.
Indeed, after gluing, the diagram (b)
has the new line containing
$q^{\mu_1,...,\mu_n}$ and, from the symmetry $q \leftrightarrow - q$,
the result for the diagram is zero for odd $n$ values.

The contribution of the diagram (a) is not zero but it is exactly cancelled
at odd $n$ values
by the contribution of
its charge conjugated diagram. Indeed, the charge conjugated diagram
coincides with the diagram (a) on Fig.~\ref{fig1} excepting gluons which propagate
to opposite direction. Then, after gluing, the sum of the diagram (a)
and  its charge conjugated diagram contains the gluon propagator with
numerator $\sim (1+(-1)^n) q^{\mu_1,...,\mu_n}$ and it is zero for odd $n$
values.

Thus, the analytic continuation is an important operation for DIS structure
function because, using the procedure \cite{KaKo,KaKoTMF}
 we calculate really another quantity: the forward scattering amplitude,
and later we apply the obtained results for moments of the DIS structure
functions. When, someone calculates a process directly, the
analytic continuation is not so important property. For example, in the
calculation of Feynman integrals with massive propagators it is convenient
to use the back mass expansion
\cite{FleKoVe}, where the
coefficients contain the considered nested sums. For the expansion, it is
possible to use both: the original nested sums and/or its analytic
continuation. The results do not depend
of a concrete choice.

\section{ Procedure of analytic continuation }

 Here we would like to consider the analytic continuation
procedure for the nested sums $S_{a,b, ...}(n)$ Eq.~(\ref{I0ad}) and
$S_{-a,b, ...}(n)$ Eq.~(\ref{I0}).
We will follow to the studies of Refs.~\cite{KaKo,Ko94}.\\

{\bf 1.} Consider the procedure of analytic continuation for
the sums $S_a(n)$ and $S_{a,b, ...}(n)$. Their form (\ref{I0ad})
is very convenient for the integer $n$ values and we should find
representations which are useful for real/complex values of their argument.\\

{\bf 1a.} As the first example, let us to study the well known
function:
$$ S_a(n)~=~\sum^n_{j=1} \frac{1}{j^a}  $$
for real and/or complex $n$ values.

Considering firstly the case $a \geq 2$, we have
\begin{eqnarray}
 S_a(n)~=~ \biggl[ \sum^{\infty }_{j=1}
-\sum^{\infty }_{j=n} \biggr]
\frac{1}{j^a} ~=~ S_a(\infty)
- \sum^{\infty }_{l=0} \frac{1}{(l+n+1)^a}
~\equiv~ S_a(\infty)
- \Psi_a(n+1),
\label{A1}
\end{eqnarray}
where
\begin{eqnarray}
S_a(\infty) ~=~ \zeta (a),~~~
\Psi_a(n+1) ~=~ \frac{(-1)^a}{(a-1)!} \Psi^{(a-1)}(n+1), \label{A2}
\end{eqnarray}
$\zeta(a)$ is the Riemann zeta functions
and $\Psi ^{(a)}(n)$ is $a$-time derivative of
the Euler $\Psi $-function, which is well known for any real and/or
complex values of $n$.

For $a=1$ case the Eq. (\ref{A1}) transforms to
\begin{eqnarray}
 S_1(n)~=~ \Psi(n+1) - \Psi(1)\,.
\label{A3}
\end{eqnarray}

{\it
One can see that the basic step  of analytic continuation
is moving
the  variable $n$ from the upper limit of the sum
to the sum argument.}
It is the basic step of our consideration here and below.\\

{\bf 1b.} Let us to continue with the function
$$ S_{a,b, ...}(n)~=~\sum^n_{j=1} \frac{1}{j^a} S_{b, ...}(j)\,.  $$

Repeating above analysis, we have got
\begin{eqnarray}
 S_{a,b, ...}(n)&=& \biggl[ \sum^{\infty }_{j=1}
-\sum^{\infty }_{j=n} \biggr]
\frac{1}{j^a}  S_{b, ...}(j) ~=~ S_{a,b, ...}(\infty)
- \sum^{\infty }_{l=0} \frac{1}{(l+n+1)^a}  S_{b, ...}(l+n+1)
\nonumber \\
&=&
S_{a,b, ...}(\infty) - \Psi_{a,b,... }(n+1), \label{A4}
\end{eqnarray}
where the function
$$
\Psi_{a,b,... }(n+1) ~=~ \sum^{\infty }_{l=0} \frac{1}{(l+n+1)^a}
 S_{b, ...}(l+n+1)
$$
is also defined for any real and/or complex
$n$ values.

From definition (\ref{I0ad}) we have that
\begin{eqnarray}
\hspace*{-7mm}S_{a,b}(\infty) &=& \zeta(a,b) +  \zeta(a+b), \nonumber \\
\hspace*{-7mm}S_{a,b,c}(\infty) &=& \zeta(a,b,c) +  \zeta(a+b,c) + \zeta(a,b+c) +
\zeta(a+b+c), \nonumber \\
\hspace*{-7mm}S_{a,b,c,d}(\infty) &=& \zeta(a,b,c,d) +  \zeta(a+b,c,d) + \zeta(a,b+c,d) +
\zeta(a,b,c+d) \nonumber \\
&+& \zeta(a+b+c,d) +  \zeta(a+b,c+d) + \zeta(a,b+c+d) +
\zeta(a+b+c+d), \label{Z1}
\end{eqnarray}
where $ \zeta(a,b,c,d, ... )$ are Eulier-Zagier alternative sums
\cite{Borwein}.

Note that sometimes (see, for example, \cite{FleKoVe}) the another
definition of the nested sums $\tilde S_{\pm a,b, ... }(n)$
 is used,
where
\begin{eqnarray}
\tilde S_{\pm a,b, ...}(n)~=~  \sum^n_{k=1} \frac{(\pm 1)^k}{k^a} \,
\tilde S_{b, ...}(k-1),~~
\tilde S_{\pm b}(n)~=~ S_{\pm b}(n)~=~ \sum^n_{k=1} \frac{(\pm 1)^k}{k^b}
 \label{I0adc}
\end{eqnarray}
and
\begin{eqnarray}
\tilde S_{\pm a,b, ... }(\infty) &=& \zeta(\pm a,b, ...),
 \label{Z2}
\end{eqnarray}
where $ \zeta(\pm a,\pm b,\pm c,\pm d, ... )$ are Eulier-Zagier
nonalternative sums \cite{Borwein}, when there is at least one negative
argument.

The Eq.~(\ref{Z1}) comes from the one (\ref{Z2}) and the relation between
the nested sums $\tilde S_{\pm a,b, ... }(n)$ and $S_{\pm a,b, ... }(n)$:
\begin{eqnarray}
S_{\pm a,b}(n) &=& \tilde S_{\pm a,b}(n) + \tilde S_{\pm (a+b)}(n),\nonumber \\
S_{\pm a,b,c}(n) &=& \tilde S_{\pm a,b,c}(n) + \tilde S_{\pm (a+b),c}(n)
+ \tilde S_{\pm a,b+c}(n) + \tilde S_{\pm (a+b+c)}(n), \nonumber \\
S_{\pm a,b,c,d}(n)
 &=& \tilde S_{\pm a,b,c,d}(n) +
\tilde S_{\pm (a+b),c,d}(n)
+ \tilde S_{\pm a,b+c,d}(n) + \tilde S_{\pm a,b,c+d}(n) \nonumber \\
&+& \tilde S_{\pm (a+b+c),d}(n) + \tilde S_{\pm (a+b),c+d}(n)
+ \tilde S_{\pm a,b+c+d}(n) + \tilde S_{\pm (a+b+c+d)}(n).
\label{Z3}
\end{eqnarray}

\vskip 0.5cm

{\bf 2.} Now we consider the procedure of analytic continuation for the sums
$S_{-a}(n)$ and $ S_{-a,b, ...}(n)$, which come in consideration of the
non-planar Feynman diagrams of forward scattering
(see, for example, the Eqs. (\ref{FI1}) and (\ref{FI2}), the diagram (b)
in  Fig.~\ref{fig1}
and discussions in Section 2).\\

\begin{figure}
\begin{center}
\epsfig{figure=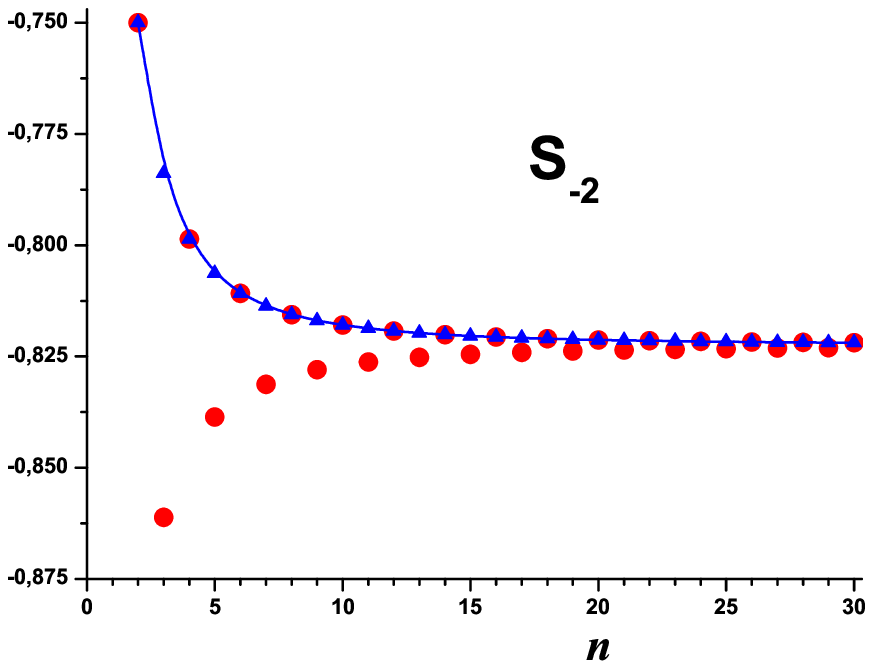,width=8cm}\epsfig{figure=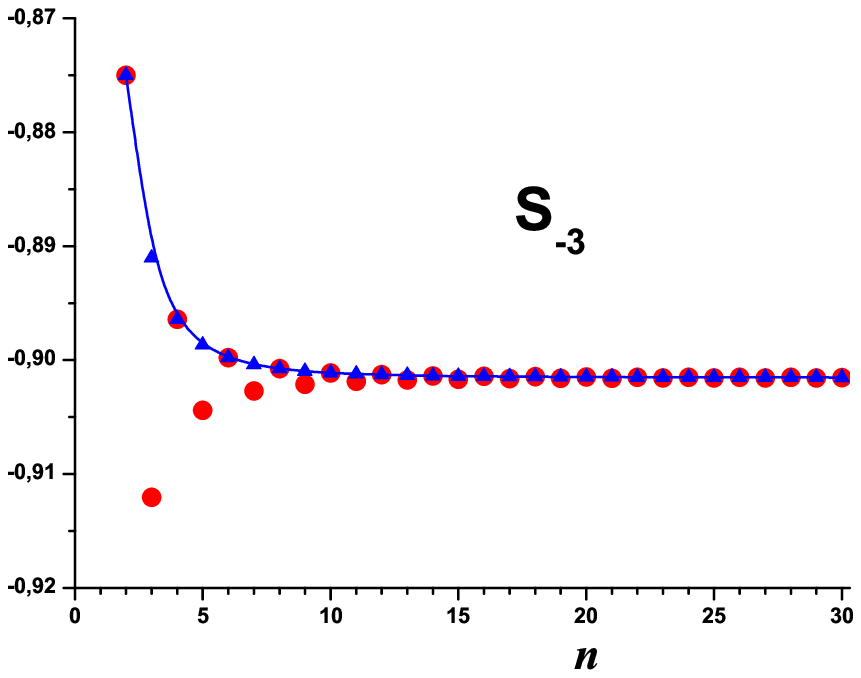,width=8cm}
\end{center}
\caption{The circles are represented the sums $S_{-2}(n)$ and $S_{-3}(n)$.
The triangles show results for $\overline S_{-2}^+(n)$ and
$\overline S_{-3}^+(n)$.}
\label{s-a}
\end{figure}

{\bf 2a.} Let us firstly to consider the functions
\footnote{ In \cite{KaKo,KaKoPl} the functions $ K_a(n)=- S_{-a}(n),~
Q(n)= K_{2,1}(n),~ K_{a,b}(n)=- S_{-a,b}(n)$ have been introduced and
their analytic continuation has been studied.}

$$ S_{-a}(n)~=~\sum^n_{j=1} \frac{(-1)^{j}}{j^a} ~=~
  -1+\frac{1}{2^a}-\frac{1}{3^a}+\frac{1}{4^a}-\frac{1}{5^a}+ ... ~,  $$
which have the smooth behavior only for even or odd parts but not for
all integer $n$ values, because there is a term $(-1)^{j}$. Indeed, we have
the two {\em different} functions for even and odd $n$ values
(see Fig.~\ref{s-a}). Thus, we
should determine firstly these two different functions
for all integer $n$ values and later
for real and complex ones. The functions, which have been analytically
continuated from even and odd $n$ values,
will be marked as $\overline S^+_{-a}(n)$ and $\overline S^-_{-a}(n)$,
respectively.

Let us to start with
$\overline S^+_{-a}(n)$, which should be
determine at its odd $n$ values. The consideration of
$\overline S^-_{-a}(n)$ will be done in the following subsection.

By analogy with the subsection {\bf 1a}, we have got
\begin{eqnarray}
 S_{-a}(n)&=&  \biggl[ \sum^{\infty }_{j=1}
- \sum^{\infty }_{j=n} \biggr]
\frac{(-1)^j}{j^a} ~=~ S_{-a}(\infty)
- \sum^{\infty }_{l=0} \frac{(-1)^{l+n+1}}{(l+n+1)^a}
\nonumber \\
&\equiv & S_{-a}(\infty)
- (-1)^n \Psi_{-a}(n+1),
\label{A5}
\end{eqnarray}
where
\begin{eqnarray}
S_{-1}(\infty) &=& \ln 2,~~
S_{-a}(\infty) ~=~  \zeta (-a)  ~=~ \left( 2^{1-a}-1 \right) \zeta (a),
\nonumber \\
\Psi_{-a}(n+1) &=& \sum^{\infty }_{l=0} \frac{(-1)^{l+1}}{(l+n+1)^a}
~=~\frac{(-1)^a}{(a-1)!} \beta^{(a-1)}(n+1) \label{A6}
\end{eqnarray}
and
 $\beta ^{(a)}(n)$ is $a$-time derivative of
the $\beta (z) $-function:
$$ \beta (z) = \frac{1}{2} \left[ \Psi \left(\frac{1+z}{2} \right) -
 \Psi \left(\frac{z}{2} \right) \right],~~~
\beta^{(a)} (z) = \frac{1}{2^{(a+1)}} \left[
\Psi^{(a)} \left(\frac{1+z}{2} \right) -
 \Psi^{(a)} \left(\frac{z}{2} \right) \right],
$$
which are
well known for any real and/or complex values of its argument.

It is clearly seen that the
nonsmooth $n$-dependence
of the  $S_{-a}(n)$ function is
indicated as the term $(-1)^n$ in front of the smooth function
$\Psi_{-a}(n+1)$.
Then, the sum
$$ (-1)^n S_{a}(n)~-~ (-1)^n  S_{-a}(\infty) = -  \Psi_{-a}(n+1) $$
is smooth on $n$, and, thus, the function
\begin{eqnarray}
 \overline S^+_{-a}(n) ~=~ (-1)^n S_{-a}(n)~+~ (1- (-1)^n)
S_{-a}(\infty)\label{A7}
\end{eqnarray}
is our needed result, because it is smooth on $n$ and coincides with the
initial one $S_{-a}(n)$ for even $n$.

The results are presented on Fig.~\ref{s-a}, where the functions
$\overline S^+_{-2}(n)$ and $\overline S^+_{-3}(n)$ demonstrate their
smooth $n$ behavior.

Note that
\begin{eqnarray}
 \overline S^+_{-a}(n) ~=~   S_{-a}(\infty)  -  \Psi_{-a}(n+1) \label{A8}
\end{eqnarray}
and, thus, it can be applied for real and/or complex $n$ values.\\

{\bf 2b.} By analogy with
the previous subsection it is possible to show,
that the
continuation of the function  $S_{-a}(n)$
from the odd $n$ values to all integer $n$ ones leads to the new
function $\overline S^-_{-a}(n)$, which
can be obtained replacing the factor $(-1)^n$ to the one
$(-1)^{n+1}$ in the r.h.s. of Eqs.~(\ref{A7}) and (\ref{A8}), i.e.
\begin{eqnarray}
 \overline S^-_{-a}(n) ~=~ (-1)^{n+1} S_{-a}(n)~+~ (1- (-1)^{n+1})
S_{-a}(\infty) ~=~ 2 S_{-a}(\infty) -  \overline S^+_{-a}(n).
\label{A9}
\end{eqnarray}

Note that the analytic continuation to the real and/or complex
$n$-values gives
\begin{eqnarray}
 \overline S^-_{-a}(n) ~=~   S_{-a}(\infty)  +  \Psi_{-a}(n+1)
\label{A8.1}
\end{eqnarray}
and, thus, the analytic continuated function $\overline S^-_{-a}(n)$
 can be defined for real and/or complex $n$ values.\\

{\bf 2c.} By analogy with the above analysis for $S_{-a}(n)$
we able to consider the function $S_{-a, b, ... }(n)$.
We have got for its analytic continuation $ \overline S^+_{-a,b,
  ...}(n)$ as
\begin{eqnarray} \overline S^+_{-a,b, ...}(n) ~=~ (-1)^n S_{-a,b, ... }(n)
~+~ (1- (-1)^n)  S_{-a, b, ... }(\infty),\label{A9.a}
\end{eqnarray}
which is defined for  real and/or complex $n$ values, because
\begin{eqnarray}
 \overline S^+_{-a, b, ... }(n) ~=~   S_{-a, b, ... }(\infty) -
\Psi_{-a, b, ... }(n+1),  \label{A10}
\end{eqnarray}
where
\begin{eqnarray}
 \Psi_{-a, b, ... }(n+1) = \sum^{\infty }_{l=0} \frac{(-1)^{l+1}}
{(l+n+1)^a} S_{b, ... }(l+n+1).\label{A11}
\end{eqnarray}

In agreement with the analysis in the subsection {\bf 2b}
the analytic continuation $\overline S^-_{-a,b, ...}(n) $
can be obtained similar to previous results: it is coincides
with the Eq.~(\ref{A9.a}) with the replacement $(-1)^n \to
(-1)^{n+1}$, i.e.
\begin{eqnarray}
 \overline S^-_{-a,b, ...}(n) &=& (-1)^{n+1} S_{-a,b, ... }(n)
~+~ (1- (-1)^{n+1})  S_{-a, b, ... }(\infty),\nonumber \\
&=&2  S_{-a, b, ... }(\infty) - \overline S^+_{-a,b, ...}(n)
\label{A9.1}
\end{eqnarray}
and can be defined for  real and/or complex $n$ values.

Indeed, for the analytic continuations $\overline S^{\pm}_{-a,b, ...}(n) $
from even and odd $n$ values, respectively, to real and/or complex $n$
values, we have
\begin{eqnarray}
 \overline S^{\pm}_{-a, b, ... }(n) ~=~   S_{-a, b, ... }(\infty) \mp
\Psi_{-a, b, ... }(n+1).  \label{A10.1}
\end{eqnarray}

Note that the functions $\overline S^+_{-a}(n) $ and
$\overline S^-_{-a}(n) $ and also $\overline S^+_{-a,b, ...}(n) $ and
$\overline S^-_{-a,b, ...}(n) $ are not independent each other
(see Eqs. (\ref{A8.1}) and (\ref{A9.1})).

From definition (\ref{I0}) we have that
\begin{eqnarray} \z
S_{-a,b}(\infty) ~=~ \zeta(-a,b) +  \zeta(-(a+b)), \nonumber \\
\z
S_{-a,b,c}(\infty) ~=~ \zeta(-a,b,c) +  \zeta(-(a+b),c) + \zeta(-a,b+c) +
\zeta(-(a+b+c)), \nonumber \\
\z
S_{-a,b,c,d}(\infty) ~=~ \zeta(-a,b,c,d) +  \zeta(-(a+b),c,d) +
\zeta(-a,b+c,d) +  \zeta(-a,b,c+d)  \nonumber \\
\z \vspace{1cm}
+ \zeta(-(a+b+c),d) +  \zeta(-(a+b),c+d) + \zeta(-a,b+c+d) +
\zeta(-(a+b+c+d)).
\label{Z4}
\end{eqnarray}

\vskip 0.5cm

{\bf 2d.} The functions $ S_{-a}(n)$ and $ S_{-a, b, ... }(n)$ are
related to the other popular
ones (see \cite{Ynd,LoYn,LoYn1}):
\begin{eqnarray} \z
 \tilde  S(n) ~=~ S_{-2,1}(n), \nonumber \\  \z
 S_l^{'}(\frac{1}{2} n) ~\equiv ~ S_l(\mbox{integer part of}
\, \, \frac{n}{2}) ~=~ 2^{l-1} \Bigl[S_l(n)+S_{-l}(n) \Bigr].  \nonumber
\end{eqnarray}

In agreement with the previous studies,
these functions can be continuated from even to all integer $n$ values.
They have the following form (see the second paper in \cite{Krivo})
\begin{eqnarray} \z
 \overline S^{'+}_2(\frac{1}{2} n) ~=~
2S_2(n)+2  \overline  S^+_{-2}(n) ~=~
(-1)^n S^{'}_2(n)+(1-(-1)^n) \Bigl[2S_2(n)- \zeta (2) \Bigr], \nonumber \\
\z
 \overline S^{'+}_3(\frac{1}{2} n) ~=~
4S_2(n)+4 \overline  S^+_{-2}(n) ~=~
(-1)^n S^{'}_3(n)+(1-(-1)^n) \Bigl[4S_2(n)- 3\zeta (3) \Bigr], \nonumber \\
\z  \overline {\tilde S}^{+}( n) ~=~
 \overline S_{-2,1}^+ (n)  ~=~
(-1)^n \tilde S(n)- (1-(-1)^n) \frac{5}{8}\zeta(3). \label{6}
\end{eqnarray}

The continuation from odd to all integer $n$ values  can be obtained by
the replacement $(-1)^n \to (-1)^{n+1}$ in the r.h.s. of above equations.

The Eq.~(\ref{6}) and the replacement $(-1)^n \to (-1)^{n+1}$ in their
r.h.s. correspond, respectively, to the ``$+$'' and ``$-$'' prescriptions
(\ref{A7}), (\ref{A9}), (\ref{A9.a}) and (\ref{A9.1}).

\section{ More complicated cases }

At the NLO approximations only the nonsmooth functions  $S_{-a}(n)$ and
$S_{-a, b, ... }(n)$ contribute to the anomalous dimensions \cite{LoYn,LoYn1}
and the longitudinal coefficient functions \cite{KaKoPl}. At NNLO level
the  new functions
$$S_{a, -b, c, ... }(n), ~ S_{-a, -b, c, ... }(n),~
S_{a, -b, -c, ... }(n),~ S_{-a, b, -c, ... }(n),~ S_{a, b, -c, ... }(n)$$
start to play a role. In principle, the results
of \cite{MVVns,MVVsi} can be represented in the form where only the one new function
$S_{a, -b, c, ... }(n)$ contributes.

However, the original form of representations of the NNLO anomalous dimensions
in \cite{MVVns,MVVsi} seems to be most compact one
and, so, it is better to keep it. In
this
case we should find the
analytic continuations for all above sums.

Moreover, we consider also the sum $S_{-a, -b, -c, ... }(n)$, which
does not contribute to the NNLO corrections to the anomalous dimensions.
However, it can contribute to future results \cite{MVVfuture} for the
coefficient functions at NNLO level.

The derivation of the formulae is done in Appendix A. The final results
for the analytic continuation from even $n$-values to integer ones has
the form
\begin{eqnarray}
\overline S^+_{-a, -b, ... }(n) &=& S_{-a, -b, ... }(n) ~+~
  (1-(-1)^n) S_{-b, ... }(\infty) \, \biggl[S_{-a}(\infty)-S_{-a}(n)\biggr],
\label{BB1} \\
\overline S^+_{a, -b, ... }(n) &=&
(-1)^n S_{a, -b, ... }(n) ~+~
(1-(-1)^n)  \biggl[  S_{a,-b, ... }(\infty) \nonumber \\
&&-  S_{-b, ... }(\infty)
\Bigl( S_{a}(\infty)-S_{a}(n) \Bigr) \biggr],
\label{BB2}\\
\overline S^+_{a, -b,-c, ... }(n) &=& S_{a, -b,-c, ... }(n) ~+~
 (1-(-1)^n)   S_{-c, ... }(\infty)  \biggl[S_{a,-b}(\infty)-S_{a,-b}(n)
 \nonumber \\
&&-S_{-b}(\infty) \Bigr( S_{a}(\infty)-S_{a}(n)\Bigr)\biggr],
\label{BB3}\\
\overline S^+_{-a, b,-c, ... }(n) &=& S_{-a, b,-c, ... }(n) ~+~
(1-(-1)^n)
\Biggl[ \biggl( S_{b,-c, ... }(\infty) -
S_{b}(\infty) S_{-c, ... }(\infty)  \biggr)
 \nonumber \\
&&\times\Bigl(S_{-a}(\infty)-S_{-a}(n)\Bigr)
~+~ S_{-c, ... }(\infty)
\biggr(S_{-a,b}(\infty)-S_{-a,b}(n) \biggr) \Biggr],
\label{BB4}\\
\overline S^+_{a, b,-c, ... }(n) &=& (-1)^n S_{a, b,-c, ... }(n)
~+~
(1-(-1)^n)
\Biggl[ S_{a, b,-c, ... }(\infty) \nonumber  \\
&&-
\biggl( S_{b,-c, ... }(\infty) -  S_{b}(\infty) S_{-c, ... }(\infty)  \biggr)
\Bigl(S_{a}(\infty)-S_{a}(n)\Bigr)
\nonumber \\&&- S_{-c, ... }(\infty)
\biggr(S_{a,b}(\infty)-S_{a,b}(n) \biggr) \Biggr],
\label{BB5}\\
\overline S^+_{-a, -b,-c, ... }(n) &=& (-1)^n S_{-a, -b,-c, ... }(n)
~+~
(1-(-1)^n)
\Biggl[ S_{-a, -b,-c, ... }(\infty) \nonumber  \\
&& \hskip -1.5cm
- S_{-c, ... }(\infty)
\biggl( S_{-a,-b}(\infty) - S_{-a,-b}(n) -  S_{-b}(\infty)
\Bigl(S_{-a}(\infty)-S_{-a}(n)\Bigr)
\biggr) \Biggr].
\label{BB6}
\end{eqnarray}

We can see that the formulae are similar to ones presented in the
previous section but they have little more complicated form.

As it was before, the analytic continuation from odd $n$ values to the
integer ones can be done by replacement all terms $(-1)^n$ by
ones $(-1)^{n+1}$, i.e.
\begin{eqnarray}
\overline S^-_{-a, -b, ... }(n) &=& S_{-a, -b, ... }(n) ~+~
  (1-(-1)^{n+1}) S_{-b, ... }(\infty) \,
\biggl[S_{-a}(\infty)-S_{-a}(n)\biggr],
\label{BB1-} \\
\overline S^-_{a, -b, ... }(n) &=&
(-1)^{n+1} S_{a, -b, ... }(n) ~+~
(1-(-1)^{n+1})  \biggl[  S_{a,-b, ... }(\infty) \nonumber \\
&&-  S_{-b, ... }(\infty)
\Bigl( S_{a}(\infty)-S_{a}(n) \Bigr) \biggr],
\label{BB2-}\\
\overline S^-_{a, -b,-c, ... }(n) &=& S_{a, -b,-c, ... }(n) ~+~
 (1-(-1)^{n+1})   S_{-c, ... }(\infty)  \biggl[S_{a,-b}(\infty)-S_{a,-b}(n)
 \nonumber \\
&&-S_{-b}(\infty) \Bigr( S_{a}(\infty)-S_{a}(n)\Bigr)\biggr],
\label{BB3-}\\
\overline S^-_{-a, b,-c, ... }(n) &=& S_{-a, b,-c, ... }(n) ~+~
(1-(-1)^{n+1})
\Biggl[ \biggl( S_{b,-c, ... }(\infty) -
S_{b}(\infty) S_{-c, ... }(\infty)  \biggr)
 \nonumber \\
&& \times\Bigl(S_{-a}(\infty)-S_{-a}(n)\Bigr)
~+~ S_{-c, ... }(\infty)
\biggr(S_{-a,b}(\infty)-S_{-a,b}(n) \biggr) \Biggr],
\label{BB4-}\\
\overline S^-_{a, b,-c, ... }(n) &=& (-1)^{n+1} S_{a, b,-c, ... }(n)
~+~
(1-(-1)^{n+1})
\Biggl[ S_{a, b,-c, ... }(\infty) \nonumber  \\
&&-
\biggl( S_{b,-c, ... }(\infty) -  S_{b}(\infty) S_{-c, ... }(\infty)  \biggr)
\Bigl(S_{a}(\infty)-S_{a}(n)\Bigr)
\nonumber \\&&- S_{-c, ... }(\infty)
\biggr(S_{a,b}(\infty)-S_{a,b}(n) \biggr) \Biggr],
\label{BB5-}\\
\overline S^-_{-a, -b,-c, ... }(n) &=& (-1)^{n+1} S_{-a, -b,-c, ... }(n)
~+~
(1-(-1)^{n+1})
\Biggl[ S_{-a, -b,-c, ... }(\infty) \nonumber  \\
&&  \hskip -1.5cm
- S_{-c, ... }(\infty)
\biggl( S_{-a,-b}(\infty) - S_{-a,-b}(n) -  S_{-b}(\infty)
\Bigl(S_{-a}(\infty)-S_{-a}(n)\Bigr)
\biggr) \Biggr].
\label{BB6-}
\end{eqnarray}

As it was shown in the previous section, the  ``$+$'' and ``$-$'' forms of
analytic continuations are not independent from each other. Indeed, we have
\begin{eqnarray}
\overline S^-_{-a, -b, ... }(n) &=&  \overline S^+_{-a, -b, ... }(n)
\ +\ 2\,(-1)^n\,  S_{-b, ... } (\infty)
\biggl[S_{-a}(\infty)-S_{-a}(n)\biggr] \nonumber \\
&=&  \overline S^+_{-a, -b, ... }(n)
\ +\ 2\, S_{-b, ... } (\infty)
\biggl[S_{-a}(\infty)- \overline S^+_{-a}(n)\biggr],
\label{BB1-+} \\
\overline S^-_{a, -b, ... }(n) &=&
2\, \biggl[  S_{a,-b, ... }(\infty) -  S_{-b, ... }(\infty)
\Bigl( S_{a}(\infty)-S_{a}(n) \Bigr) \biggr]
~-~ \overline S^+_{a, -b, ... }(n),
\label{BB2-+}\\
\overline S^-_{a, -b,-c, ... }(n) &=& \overline S^+_{a, -b,-c, ... }(n)
\ +\ 2\,(-1)^n\, S_{-c, ... }(\infty)  \biggl[S_{a,-b}(\infty)-S_{a,-b}(n)
 \nonumber \\
&&-S_{-b}(\infty) \Bigr( S_{a}(\infty)-S_{a}(n)\Bigr)\biggr]
~=~ \overline S^+_{a, -b,-c, ... }(n) \ +\ 2\, S_{-c, ... }(\infty)
\nonumber \\
&&\times  \biggl[S_{a,-b}(\infty)-
\overline S^+_{a,-b}(n) -S_{-b}(\infty) \Bigr( S_{a}(\infty)-S_{a}(n)\Bigr)
\biggr],
\label{BB3-+}\\
\overline S^-_{-a, b,-c, ... }(n) &=& \overline S^+_{-a, b,-c, ... }(n)
\ +\ 2\,(-1)^n\,  \Biggl[ \biggl( S_{b,-c, ... }(\infty) -
S_{b}(\infty) S_{-c, ... }(\infty)  \biggr)
 \nonumber \\
&&\times \Bigl(S_{-a}(\infty)-S_{-a}(n)\Bigr)
~+~ S_{-c, ... }(\infty)
\biggr(S_{-a,b}(\infty)-S_{-a,b}(n) \biggr) \Biggr] \nonumber \\
&=& \overline S^+_{-a, b,-c, ... }(n)
\ +\ 2\,  \Biggl[ \biggl( S_{b,-c, ... }(\infty) -
S_{b}(\infty) S_{-c, ... }(\infty)  \biggr)
 \nonumber \\
&&\times \Bigl(S_{-a}(\infty)- \overline  S^+_{-a}(n)\Bigr)
~+~ S_{-c, ... }(\infty)
\biggr(S_{-a,b}(\infty)- \overline S^+_{-a,b}(n) \biggr) \Biggr],
\label{BB4-+}\\
\overline S^-_{a, b,-c, ... }(n)
&=& 2\,\Biggl[ S_{a, b,-c, ... }(\infty)
~-~
\biggl( S_{b,-c, ... }(\infty) -  S_{b}(\infty) S_{-c, ... }(\infty)  \biggr)
\Bigl(S_{a}(\infty)-S_{a}(n)\Bigr)
\nonumber \\&&- S_{-c, ... }(\infty)
\biggr(S_{a,b}(\infty)-S_{a,b}(n) \biggr) \Biggr]
-\overline S^+_{a, b,-c, ... }(n),
\label{BB5-+}\\
\overline S^-_{-a, -b,-c, ... }(n) &=&
2 \,
\Biggl[ S_{-a, -b,-c, ... }(\infty) - S_{-c, ... }(\infty)
\biggl( S_{-a,-b}(\infty) - S_{-a,-b}(n)  \nonumber  \\
&&-  S_{-b}(\infty)
\Bigl(S_{-a}(\infty)-S_{-a}(n)\Bigr)
\biggr) \Biggr] - \overline S^+_{-a, -b,-c, ... }(n) \nonumber \\
&=&
2 \,
\Biggl[ S_{-a, -b,-c, ... }(\infty) - S_{-c, ... }(\infty)
\biggl( S_{-a,-b}(\infty) -\overline  S^+_{-a,-b}(n)  \nonumber  \\
&&-  S_{-b}(\infty)
\Bigl(S_{-a}(\infty)-\overline S^+_{-a}(n)\Bigr)
\biggr) \Biggr] - \overline S^+_{-a, -b,-c, ... }(n).
\label{BB6-+}
\end{eqnarray}

Similar to Eq. (\ref{A9.1}), the relations between the  ``$+$'' and
``$-$'' forms of analytic continuations have smooth $n$-dependence.

The analytic continuation to real and/or complex $n$ values can be easy
obtained from above formulae. It has the form (see Appendix B for details):
\begin{eqnarray}
\overline S_{-a,-b, ... }^{+}(n) &=& S_{-a,-b, ... }(\infty) -
\Psi_{a,-b, ... }(n+1) \nonumber \\
&+& S_{-b, ... }(\infty) \Bigl[ \Psi_{a}(n+1)- \Psi_{-a}(n+1)\Bigr],
\label{BBB1}\\
 \overline S_{a,-b, ... }^{+}(n) &=& S_{a,-b, ... }(\infty) -
\Psi_{-a,-b, ... }(n+1) \nonumber \\
&+& S_{-b, ... }(\infty) \Bigl[ \Psi_{-a}(n+1)- \Psi_{a}(n+1)\Bigr],
\label{BBB2}\\
 \overline S_{a,-b,-c, ... }^{+}(n) &=& S_{a,-b,-c, ... }(\infty) -
\Psi_{a,-b,-c, ... }(n+1) \nonumber \\
&+& S_{-c, ... }(\infty) \biggl[ \Psi_{-a,-b}(n+1)- \Psi_{a,-b}(n+1)
\nonumber \\
&-&
S_{-b}(\infty) \Bigl( \Psi_{a}(n+1)- \Psi_{-a}(n+1)
\Bigr) \biggr],
\label{BBB3}\\
 \overline S_{-a,b,-c, ... }^{+}(n) &=& S_{-a,b,-c, ... }(\infty) -
\Psi_{a,b,-c, ... }(n+1) \nonumber \\
&+& \Bigl[S_{b,-c, ... }(\infty) -S_{b}(\infty) S_{-c, ... }(\infty) \Bigr]
\Bigl( \Psi_{a}(n+1)- \Psi_{-a}(n+1)\Bigr)
\nonumber \\
&+& S_{-c, ... }(\infty) \Bigl( \Psi_{a,b}(n+1)- \Psi_{-a,b}(n+1)\Bigr),
\label{BBB4}\\
 \overline S_{a,b,-c, ... }^{+}(n) &=& S_{a,b,-c, ... }(\infty) -
\Psi_{-a,b,-c, ... }(n+1) \nonumber \\
&+&
\Bigl[S_{b,-c, ... }(\infty) -S_{b}(\infty)S_{-c, ... }(\infty)\Bigr]
\Bigl( \Psi_{-a}(n+1)- \Psi_{a}(n+1)\Bigr)
\nonumber \\
&+& S_{-c, ... }(\infty)
\Bigl( \Psi_{-a,b}(n+1)- \Psi_{a,b}(n+1)\Bigr),
\label{BBB5}\\
 \overline S_{-a,-b,-c, ... }^{+}(n) &=& S_{-a,-b,-c, ... }(\infty) -
\Psi_{-a,-b,-c, ... }(n+1) \nonumber \\
&+& S_{-c, ... }(\infty) \biggl[ \Psi_{-a,-b}(n+1)- \Psi_{a,-b}(n+1)
\nonumber \\
&+& S_{-b}(\infty) \Bigl( \Psi_{a}(n+1)- \Psi_{-a}(n+1)
\Bigr) \biggr],
\label{BBB6a}
\end{eqnarray}
where
\begin{eqnarray}
\Psi_{a,-b, ... }(n+1) &=& \sum_{l=0}^{\infty} \frac{1}{(l+n+1)^a}
\overline S_{-b,  ... }^{+}(l+n+1),
\label{BBB6}\\
\Psi_{-a,-b, ... }(n+1) &=& \sum_{l=0}^{\infty} \frac{(-1)^{l+1}}{(l+n+1)^a}
\overline S_{-b,  ... }^{+}(l+n+1),
\label{BBB7}\\
\Psi_{a,-b,-c, ... }(n+1) &=& \sum_{l=0}^{\infty} \frac{1}{(l+n+1)^a}
\overline S_{-b,-c,  ... }^{+}(l+n+1),
\label{BBB8}\\
\Psi_{a,b,-c, ... }(n+1) &=& \sum_{l=0}^{\infty} \frac{1}{(l+n+1)^a}
\overline S_{b,-c  ... }^{+}(l+n+1),
\label{BBB9}\\
\Psi_{-a,b,-c, ... }(n+1) &=& \sum_{l=0}^{\infty} \frac{(-1)^{l+1}}{(l+n+1)^a}
\overline S_{b,-c,  ... }^{+}(l+n+1),
\label{BBB10}\\
\Psi_{-a,-b,-c, ... }(n+1) &=& \sum_{l=0}^{\infty} \frac{(-1)^{l+1}}{(l+n+1)^a}
\overline S_{-b,-c,  ... }^{+}(l+n+1).
\label{BBB11}
\end{eqnarray}

So,  now the functions $\overline S_{-a,-b, ... }^{+}(n)$,
$\overline S_{a,-b, ... }^{+}(n)$,
$\overline S_{a,-b,-c, ... }^{+}(n)$, $\overline S_{-a,-b,-c, ... }^{+}(n)$,
$\overline S_{-a,b,-c, ... }^{+}(n)$ and $\overline S_{a,b,-c, ... }^{+}(n)$
are well defined for
the real and/or complex $n$ values.

The results for the analytic continuation to  real and/or complex $n$ values
of the corresponding
functions $\overline S_{-a,-b, ... }^{-}(n)$,
$\overline S_{a,-b, ... }^{-}(n)$,
$\overline S_{a,-b,-c, ... }^{-}(n)$, $\overline S_{-a,-b,-c, ... }^{-}(n)$,
$\overline S_{-a,b,-c, ... }^{-}(n)$ and $\overline S_{a,b,-c, ... }^{-}(n)$
can be found taking together
Eqs. (\ref{BB1-+})-(\ref{BB6-+}) and (\ref{BBB1})-(\ref{BBB11}).

\section{ Simple example }

As some examples we will study analytic continuation of the nonsinglet
parts of NLO and NNLO anomalous dimensions and NNLO Wilson coefficient
functions. The NLO  nonsinglet anomalous dimension
$\gamma^{(1),\pm}_{NS}(n)$ will be considered in this section and other
variables will be studied in the following one.

Here we will follow to the form of
$\gamma^{(1)\pm}_{NS}$ given in perfect Yndurain book \cite{Ynd}:
\begin{eqnarray}
\gamma^{(1),\pm}_{NS}(n) ~=~ \frac{64}{9} \Bigl[ A_1(n) ~+~ A_2(n)
~\pm~ A_3(n) \Bigr],
\label{1J}
\end{eqnarray}
where
\begin{eqnarray}
 A_1(n) &=& S_3(n) ~-~ 4 S_2(n) \Bigl[ 2S_1(n)- \frac{1}{n(n+1)}
 +\frac{21}{8} \Bigr] \nonumber \\
&+& S_1(n) \Bigl[ \frac{67}{2}+ \frac{4(2n+1)}{n^2(n+1)^2} \Bigr]
~-~ \frac{63}{16} ~-~ \frac{151n^4+260n^3+96n^2+3n+10}{4n^3(n+1)^3}
\nonumber \\
&+& f
 \Bigl[ S_2(n)-\frac{5}{3} S_1(n)+  \frac{11n^2+5n-3}{6n^2(n+1)^2}
 +\frac{1}{8} \Bigr], \label{2J}\\
A_2(n) &=&  S_{-3}(n)- 2S_{-2,1}(n)+ S_{-2}(n)
\Bigl[ 2S_1(n) - \frac{1}{n(n+1)}   \Bigr], \label{3J}\\
A_3(n) &=&  \frac{2n^2+2n+1}{2n^3(n+1)^3} \label{4J}
\end{eqnarray}
and $f$ is the number of active quarks.

Formally, the difference between the anomalous dimensions
$\gamma^{(1),+}_{NS}$ and
$\gamma^{(1),-}_{NS}$ is proportional to function $A_3(n)$.
After the analytic continuation from even and odd $n$ values for the
AD $\gamma^{(1),+}_{NS}$ and
$\gamma^{(1),-}_{NS}$, respectively, the situation changes essentially.

Indeed, in agreement with the previous section to extend the results
(\ref{1J}) to integer, real and/or complex $n$ values we can use ``$+$''
and ``$-$'' prescriptions (\ref{A7})-(\ref{Z4}) for the
anomalous dimensions  $\gamma^{(1),+}_{NS}$ and
$\gamma^{(1),-}_{NS}$, respectively,

Then, we have the analytically continuated anomalous dimensions  in the form
\begin{eqnarray}
\gamma^{(1),\pm}_{NS}(n) ~=~ \frac{64}{9} \Bigl[ A_1(n) ~+~ \overline
A_2^{\pm}(n)
~\pm~ A_3(n) \Bigr],
\label{5J}
\end{eqnarray}
where
\begin{eqnarray}
\overline A_2^{\pm}(n) ~=~  \overline S_{-3}^{\pm}(n)-
2\overline S_{-2,1}^{\pm}(n)+
\overline S_{-2}^{\pm}(n)
\Bigl[ 2S_1(n) - \frac{1}{n(n+1)}   \Bigr]. \label{6J}
\end{eqnarray}

{\bf 1.} It is useful to see
the difference between anomalous dimensions
$ \gamma^{(1)-}_{NS}$ and $ \gamma^{(1)+}_{NS}$ :
\begin{eqnarray}
\gamma^{(1)-}_{NS} ~-~  \gamma^{(1)+}_{NS} ~=~ \frac{64}{9} \Bigl[
\hat A_2(n) ~-~ 2A_3(n) \Bigr],
\label{7J}
\end{eqnarray}
where
\begin{eqnarray}
\hat A_2(n) ~=~ \overline A_2^{-}(n) - \overline A_2^{+}(n) ~=~
\hat S_{-3}(n) -2 \hat S_{-2,1}(n) - \hat S_{-2}(n) \Bigl[
2S_1(n)  - \frac{1}{n(n+1)} \Bigr]
\label{7J.a}
\end{eqnarray}
and
\begin{eqnarray}
\hat S_{-a}(n) &=& \overline S_{-a}^{-}(n) - \overline S_{-a}^{+}(n)
~=~ 2 \Bigl( S_{-a}(\infty) - \overline S_{-a}^{+}(n)\Bigr)
~=~ 2 (-1)^n \Bigl( S_{-a}(\infty) - S_{-a}(n)\Bigr),
\nonumber \\
\hat S_{-a,b}(n) &=& \overline S_{-a,b}^{-}(n) - \overline S_{-a,b}^{+}(n)
~=~ 2 \Bigl( S_{-a,b}(\infty) - \overline S_{-a,b}^{+}(n)\Bigr)
\nonumber \\
&=& 2  (-1)^n \Bigl( S_{-a,b}(\infty) - S_{-a,b}(n)\Bigr).
\label{8j}
\end{eqnarray}

For the several first $n$ values the results for anomalous dimensions
$ \gamma^{(1)-}_{NS}$ and $ \gamma^{(1)+}_{NS}$ and their difference
are give in the Table 1.
One can see, that the ratio of $(\gamma^{(1)+}_{NS}-
  \gamma^{(1)-}_{NS})/ \gamma^{(1)+}_{NS}$, which is equal to 1
at $n=1$, is very small already started with $n\geq 2$.

\begin{table}
\caption{the results for anomalous dimensions
$ \gamma^{(1)-}_{NS}$ and $ \gamma^{(1)+}_{NS}$ and their difference
at the first four even $n$ values.}
\begin{center}
\begin{tabular}{|l||l|l|l|l|}         \hline
$n$                   & 2  &  4 & 6  & 8 \\ \hline
$ \gamma^{(1)+}_{NS}$  &   77.70  &   133.25  &   164.26 &   186.68  \\ \hline
$ \gamma^{(1)+}_{NS}- \gamma^{(1)-}_{NS}$  & 1.335 $\cdot 10^{-1}$ &
4.6$\cdot 10^{-3}$  & 4$\cdot 10^{-4}$  & 7$\cdot 10^{-5}$  \\ \hline
$\frac{ \gamma^{(1)+}_{NS}- \gamma^{(1)-}_{NS}}{ \gamma^{(1)+}_{NS}}$
  &  1.87$\cdot 10^{-3}$ & 3.9$\cdot 10^{-5}$  &  3$\cdot 10^{-6}$ & 4$\cdot
  10^{-7}$
  \\ \hline
\end{tabular}
\end{center}
\end{table}

Considering  $1/n$ expansion (see \cite{LoYn,LoYn1}), which is the very good
approximation starting
with $n=4$, we have:
 \begin{eqnarray}
 \hat A_2(n) ~=~ \frac{2}{n^4} \left(1- \frac{2}{n} + \frac{13}{2n^2} \right)
+   O\left(\frac{1}{n^7}\right),~~
~A_3(n) ~=~ \frac{1}{n^4} \left(1- \frac{2}{n} + \frac{7}{2n^2} \right)
+  O\left(\frac{1}{n^7}\right)
\label{10j}
\end{eqnarray}
and, thus,
 \begin{eqnarray}
\gamma^{(1)-}_{NS} ~-~  \gamma^{(1)+}_{NS}~=~ \frac{128}{3}\frac{1}{n^6}
+  O\left(\frac{1}{n^7}\right)
\label{10j.1}
\end{eqnarray}

It is possible to show the similar property for the NNLO anomalous dimensions,
i.e.
$(\gamma^{(2)+}_{NS}(n)- \gamma^{(2)-}_{NS}(n))/ \gamma^{(2)+}_{NS}(n)\ll 1$ for $n \geq 2$.
The property was important
for fits of experimental data of $xF_3$ structure functions at NNLO
approximation (see first three papers in \cite{Kataev}).
At that time, the results for the anomalous dimension
$\gamma^{(2)-}_{NS}(n)$ have been unknown
and  it has been replaced by $\gamma^{(2)+}_{NS}(n)$.\\

{\bf 2.} It is interesting to see the values of so-called
Adler and Gottfried sum rules, $I_3^-$ and $I_2^+$, respectively,
which have the following form at the first three orders of perturbation
theory ($l=2,3$):
 \begin{eqnarray}
I_l^{\pm} &=& N_l^{\pm} \, C_l^{\pm}(\as(Q^2)) \,
A^{\pm}(\as(Q^2)), \nonumber \\
 C_l^{\pm}(\as(Q^2)) &=& 1+ \as(Q^2) \, B_{l,NS}^{(1)}(n=1)+ \as^2(Q^2) \,
B_{l,NS}^{(2),\pm}(n=1) +
O\Bigl(\as^3(Q^2)\Bigr), \nonumber \\
 A^{\pm}(\as(Q^2)) &=& 1+ \as(Q^2) d_1^{\pm} + \as^2(Q^2)
\biggl[d_2^{\pm} + \Bigl(d_1^{\pm}-b_1\Bigl)
\,d_1^{\pm}\biggr] +
O\Bigl(\as^3(Q^2)\Bigr),
\label{11j}
\end{eqnarray}
where $N_l^{\pm}$ are normalization constants, $C_l^{\pm}(\as(Q^2))$ are
coefficient functions at $n=1$,
$A^{\pm}(\as(Q^2))$ are expansions of the corresponding renormalization
exponents and $\as(Q^2)$ is QCD coupling constant. Note that
\begin{eqnarray}
N_2^{+}~=~\frac{1}{3},~~~ N_3^{-}~=~2,~~~
d_i^{\pm}~=~ \frac{\gamma^{(i),\pm}_{NS}(n=1)}{2\beta_0},~~~ b_i ~=~
\frac{\beta_i}{\beta_0}
\label{11.1j}
\end{eqnarray}
and $\beta_i$ are several first coefficient in expansion of QCD $\beta
$-function on $\as $. We put also $\gamma^{(0)}_{NS}(n=1)=0$
and use $B_{l,NS}^{(1)}= B_{l,NS}^{(1),+}= B_{l,NS}^{(1),-}$, because
only planar diagrams contribute to NLO coefficient functions and, thus, the
coefficient $B_{l,NS}^{(1)}$ has the same form at even and odd $n$ values.

Considering Eqs. (\ref{2J})-(\ref{6J}), we obtain
\begin{eqnarray}
A_1(n=1)&=& \frac{13}{16} ,~~~ A_3(n=1)~=~ \frac{5}{16} , \nonumber \\
\overline A_2^{-}(n=1) &=& - \frac{1}{2} ,~~~
 \overline A_2^{+}(n=1) ~=~ \frac{1}{2}+\zeta(3)-\frac{3}{2}\zeta(2) .
\label{11.2j}
\end{eqnarray}
and, thus, we have (in agreement with \cite{RoSa}-\cite{BroKaMa})
\begin{eqnarray}
\gamma^{-}(n=1) ~=~ 0,~~~
 \gamma^{+}(n=1) ~=~ \frac{8}{9} \, \Bigl[13+8\zeta(3)-12\zeta(2)\Bigr] .
\label{11.3j}
\end{eqnarray}

From Eqs. (\ref{11.3j}) we see that at NLO approximation
\begin{eqnarray}
\hspace*{-7mm}I_3^{-}&=& N_3^{-}~=~2, \nonumber \\
\hspace*{-7mm}I_2^{+}&=&N_2^{+}  \left(1+ \frac{\gamma^{(i),\pm}_{NS}(n=1)}{2\beta_0}
\as(Q^2) \right) = \frac{1}{3}
 \left(1+ \frac{4(13+8\zeta(3)-12\zeta(2))}{3(11-2f)}
\as(Q^2) \right),
\label{11.4j}
\end{eqnarray}
i.e. the Adler sum rule is exact and Gottfried one is violated in perturbation
theory.

Note, that the term $\zeta(2)$ cannot be obtained in calculation of the
propagator-type diagrams and, thus, it cannot contribute to functions
$T_{i,n}(Q^2)$ $(i=2,L,3)$. So, its appearance in the results for
the anomalous dimension $ \gamma^{+}(n=1)$ is exactly
the result of the analytic continuation.

\section{ Other examples }

The Ref. \cite{MoVe1}
contains the $x$- and $n$-dependencies for full set of the NLO
anomalous dimensions and NNLO coefficient functions. One can see that all
results
can be represented through the functions $S_{a,b, ...}(n\pm k)$ and
$\overline S^{+}_{-a,b, ...}(n\pm k)$ (or
$\overline S^{-}_{-a,b, ...}(n\pm k)$).
In this new representation all terms proportional to the factor $(-1)^n$
will be cancelled and
the structure of the results will be simplified.

Taking, for example, the results for the nonsinglet parts of the
NLO anomalous dimensions and NNLO coefficient functions, we have
\begin{eqnarray}
\gamma^{(1),+}_{NS}(n=1)&=&
\frac{8}{3} (C_A-2C_F) \, \Bigl[13+8\zeta(3)-12\zeta(2)\Bigr],
\label{11.5j} \\
B_{2,NS}^{(2),\pm}(n=1)&=&
(C_AC_F-2C_F^2) \,
\left[
\frac{141}{64}-\frac{21}{8}\zeta(2)+\frac{45}{8}\zeta(3) - 6 \zeta(4)
\right]
\nonumber \\
&\approx & -0.615732 ,
\label{11.6j}
\end{eqnarray}
where $C_A=N$, $C_F=(N^2-1)/(2N)$ for $SU(N)$ gauge group and $T_F=f/2$.

The result (\ref{11.5j}) is completely coincide with above one (\ref{11.3j}).
The result (\ref{11.6j}) is exactly coincides with one
from Ref.~\cite{BroKaMa}, obtained
by integration of the corresponding splitting-functions.

As it was noted already in Introduction, the NNLO corrections to the
anomalous dimensions have been recently calculated in \cite{MVVns} and
\cite{MVVsi}. The results have been done in the $x$- and $n$-spaces.
In the last case, the results have been presented only for even and for
add $n$ values, respectively, for $C$-symmetric and $C$-antisymmetric
functions.

Using the analytic continuation done before we can represent the
\cite{MVVns} and \cite{MVVsi} results in the form which is correct
for arbitrary $n$ values.

As it has been shown above, to do analytic continuation for
the results
of $\gamma^{(2),+}_{NS}(n)$  and $\gamma^{(2),-}_{NS}(n)$
 from even and odd $n$
values, respectively, we should perform the following replacement
in Eqs.~(3.5)-(3.9) of~\cite{MVVns}:
\begin{eqnarray}
 S_{-a}(n) &\to&  \overline S^{\pm}_{-a}(n), ~~~~~
S_{-a,b, ...}(n) ~\to ~ \overline S^{\pm}_{-a,b, ...}(n), \nonumber \\
 S_{-a}(n\pm 1) &\equiv& N_{\pm} S_{-a,b}(n) ~\to~
 \overline S^{\mp}_{-a}(n\pm1) ~\equiv~ N_{\pm} \overline S^{\pm}_{-a}(n),
\nonumber \\
 S_{-a,b, ...}(n\pm 1) &\equiv& N_{\pm}S_{-a,b, ...}(n) ~\to~
 \overline S^{\mp}_{-a,b, ...}(n\pm1)  ~\equiv~ N_{\pm}
 \overline S^{\pm}_{-a,b, ...}(n),
 \label{13J}
\end{eqnarray}
because $n+1$ is odd (even) if $n$ is even (odd).

In the singlet case where there are additional shifts $n \to n+m$ $(m>1)$,
the analytic continuation
should be completed by  more general formulae
\begin{eqnarray}
 S_{-a}(n\pm 2k) &\equiv& N_{\pm 2k} S_{-a}(n) ~\to~
 \overline S^{\pm}_{-a}(n\pm 2k) ~\equiv~ N_{\pm 2k}
\overline S^{\pm}_{-a}(n), \nonumber \\
S_{A,B,C, ...}(n\pm 2k) &\equiv& N_{\pm 2k} S_{A,B,C, ...}(n) ~\to~
 \overline S^{\pm}_{A,B,C, ...}(n\pm 2k) ~\equiv~ N_{\pm 2k}
 \overline S^{\pm}_{A,B,C, ...}(n), \nonumber \\
S_{-a}(n\pm 2k+1) &\equiv& N_{\pm (2k+1)} S_{-a}(n) ~\to~
 \overline S^{\mp}_{-a}(n\pm 2k+1) ~\equiv~ N_{\pm (2k+1)}
 \overline S^{\mp}_{-a}(n)  , \nonumber \\
S_{A,B,C, ...}(n\pm 2k+1) &\equiv& N_{\pm (2k+1)} S_{A,B,C, ...}(n) ~\to \nonumber \\
&&\ \to\ \overline S^{\mp}_{A,B,C, ...}(n\pm 2k+1)
\ \equiv\ N_{\pm (2k+1)}
 \overline S^{\mp}_{A,B,C, ...}(n),   \label{14J}
\end{eqnarray}
where, at least, one of indices $A$, $B$ or $C$ should be negative.

Thus, the results are correct now at arbitrary $n$ values and changed
very little to compare with original ones in \cite{MVVns} and \cite{MVVsi}.

For example, for the anomalous dimension
$\gamma^{(2),+}_{NS}(n)$ we have the following results at $n=1$:
\begin{eqnarray}
\gamma^{(2),+}_{NS}(n=1)&=& (C_F^2-C_AC_F/2) \,
\biggl\{
C_F \biggl[
290-248\zeta(2)+656\zeta(3)-1488\zeta(4)+832\zeta(5) \nonumber \\
\z+192\zeta(2)\zeta(3)
\biggr]
+ C_A \biggl[
\frac{1081}{9}+\frac{980}{3}\zeta(2)-\frac{12856}{9}\zeta(3)
+\frac{4232}{3}\zeta(4)-448\zeta(5)  \nonumber \\
\z-192\zeta(2)\zeta(3)
\biggr]
+ 2T_F \biggl[
-\frac{304}{9}-\frac{176}{3}\zeta(2)+\frac{1792}{9}\zeta(3)
+\frac{272}{3}\zeta(4)
\biggr]\biggr\}  \nonumber \\
&\approx & 161.713785 - 2.429260 \, f \, ,
\label{11.7j}
\end{eqnarray}
which is exactly coincides with one of Ref.~\cite{BroKaMa}, obtained
by integration of the corresponding splitting-functions.

\section{Conclusion}

As a conclusion we would like to stress that we presented here
the analytic continuation of the nested sums
$ N_{\pm m} S_{\pm a,\pm b,\pm c, ...}(n)$,
that is important for
$n$-space representation of the moments of the DIS structure functions.
Our results have quite compact form and change only little the original
form of the MVV representations for anomalous dimensions.

Indeed, these  nested sums contributing to the
coefficient functions and anomalous dimensions
for $C$-symmetric and $C$-antisymmetric structure functions should
be replaced, respectively, by their analytic continuations
$ N_{\pm m}  \overline S^{+}_{\pm a,\pm b,\pm c, ...}(n)$ and
$ N_{\pm m}  \overline S^{-}_{\pm a,\pm b,\pm c, ...}(n)$
(see Eq. (\ref{14J})).

We hope that the analytic continuation will be useful for presentation of
the future results for the NNLO corrections to coefficient functions.
For example, the results for the 3-loop coefficient functions of the
longitudinal structure function will be available in the nearest future
\cite{MVVfuture}:
its compact parameterizations have been already published  very recently
\cite{MVVFL}.

The analytic continuation will be important for new fits of experimental
data with help of the orthogonal polynomials. The usage of the results
allows to avoid the numerical integration of the splitting functions and to
improve the DGLAP evolution procedure in the fitting program.

The results of the analytic continuation allows also to extend to NNLO
accuracy the
analysis of small $x$ behavior of gluon density, $F_2(x,Q^2)$ and $F_L(x,Q^2)$
structure functions,
done in \cite{KoPaGlu}, \cite{Q2evo} and \cite{KoFL,KoPaFL}, respectively.

\vspace{10mm} {\large \textbf{Acknowledgments.}}\\[5mm]

We are grateful to A.L. Kataev, L.N. Lipatov and
S. Moch for useful discussions and comments.

This work was supported by the Alexander von Humboldt fellowship (A.V. K.),
 the RFBR grants 04-02-17094, RSGSS-1124.2003.2 (V.N. V.).\\


\section{ Appendix A}
\def\theequation{A\arabic{equation}}
\label{App:A}
\setcounter{equation}0

Here we derive the analytic continuation of the nested sums
$$S_{a, -b, c, ... }(n), ~ S_{-a, -b, c, ... }(n),~
S_{a, -b, -c, ... }(n),~ S_{-a, b, -c, ... }(n),~ S_{a, b, -c, ... }(n),~
S_{-a, -b, -c, ... }(n)$$
from the even $n$ values to the integer ones. The similar procedure
can be done for the analytic continuation from the odd $n$ values
and also to real and/or complex $n$ values (see Section 4 and Appendix B).

To simplify all formulae in the Appendix
we define
\begin{eqnarray}
S_{A,B,C, ...}(\infty) \equiv Z_{A,B,C, ...},
\label{B0}
\end{eqnarray}
where the symbols $A,B$ and $C$ may have positive and negative values.\\

{\bf 1}. It is better to start with the case $S_{a, -b}(n)$, where there
are only two indexes $a$ and $b$ and, respectively, there are very simple
relations between different functions. The more general case
$S_{a, -b, c, ... }(n)$ will be considered below in the subsection {\bf 3}.

Because there is a transformation
\begin{eqnarray}
S_{a, -b}(n) ~=~ S_{a}(n)S_{-b}(n) + S_{-(a+b)}(n)- S_{-b, a}(n),
\label{B1}
\end{eqnarray}
we can use for the r.h.s. of (\ref{B1}) the results of (\ref{A7}) and
  (\ref{A9.a})  obtained in ~\cite{KaKo,Ko94}.

Then, we have for analytical continuation $\overline S_{a, -b}^+(n)$
\begin{eqnarray}
\overline S^+_{a,-b}(n) ~=~ S_{a}(n)\overline S_{-b}^+(n) +
\overline S_{-(a+b)}^+(n)- \overline S_{-b, a}^+(n),
\label{B2}
\end{eqnarray}
where the functions in the r.h.s. of (\ref{B2}) are defined by Eqs.
(\ref{A7}) and (\ref{A9.a}). Using the relation  (\ref{B1}) for the
 r.h.s. of (\ref{B2}) we can easily obtain
\begin{eqnarray}
\overline S^+_{a,-b}(n) &=& (-1)^n \biggl[S_{a}(n)S_{-b}(n) +
S_{-(a+b)}(n)- S_{-b, a}(n)\biggr] \nonumber \\
&+& (1-(-1)^n) \biggl[S_{a}(n)Z_{-b} +
Z_{-(a+b)}- Z_{-b, a}\biggr]  \nonumber \\
&=& (-1)^n S_{a, -b}(n) + (1-(-1)^n)
\biggl[ Z_{a, -b} -
Z_{-b} \Bigl( Z_{a}-S_{a}(n) \Bigr)  \biggr].
\label{B3}
\end{eqnarray}

Thus, we see the additional term $\sim ( Z_{a}-S_{a}(n))$ in the
 r.h.s. of (\ref{B3}) to compare with the results of (\ref{A7}) and
  (\ref{A9.a}). \\

{\bf 2}. Consider now the sum $S_{-a, -b, ... }(n)$:

\begin{figure}
\begin{center}
\epsfig{figure=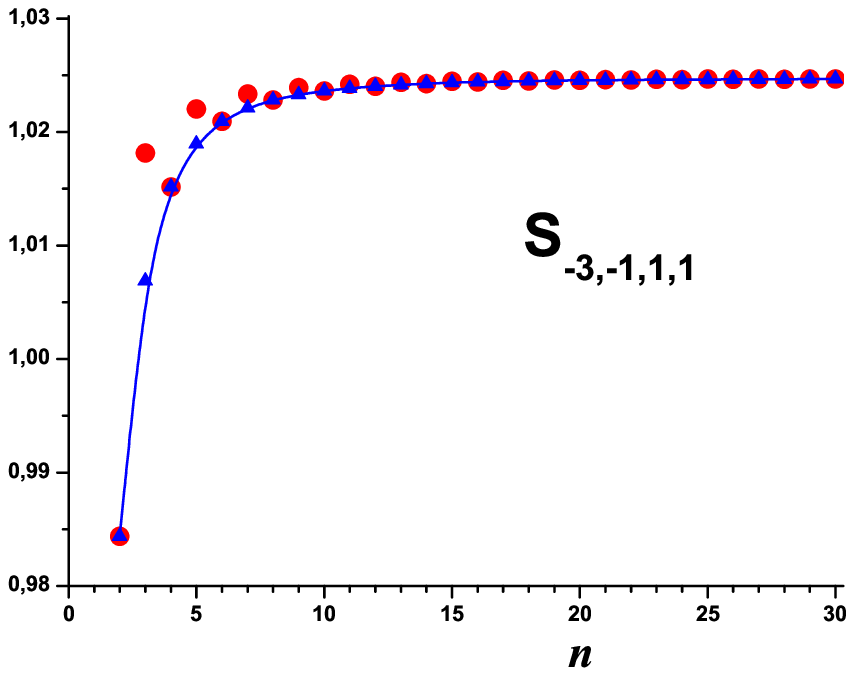,width=8cm}\epsfig{figure=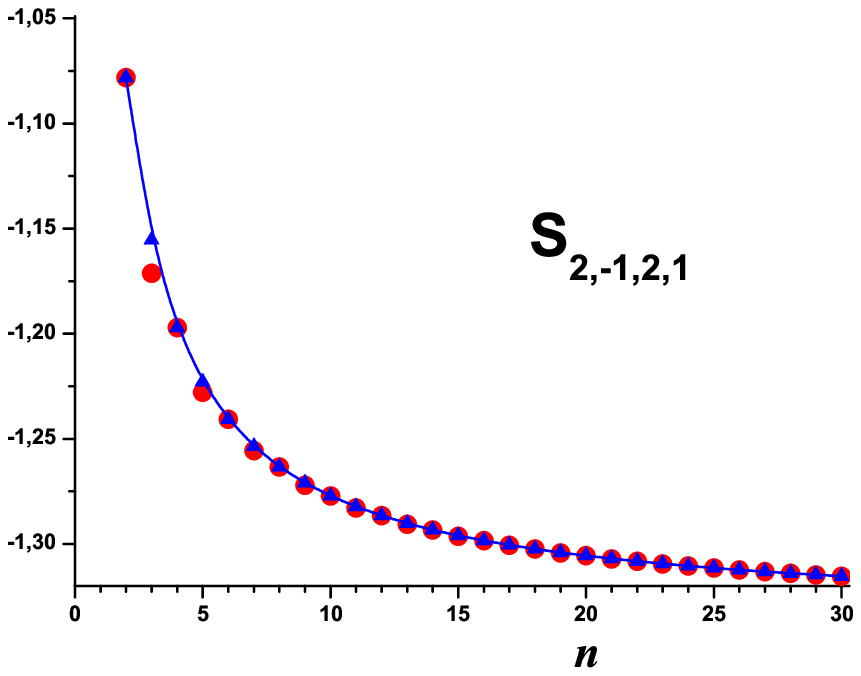,width=8cm}
\end{center}
\caption{As in Fig.~\ref{s-a} but for the sums $S_{-3, -1, 1, 1}(n)$ and $S_{2, -1, 2, 1}(n)$.}
\label{sa-bcds-a-bcd}
\end{figure}

\begin{eqnarray}
S_{-a, -b, ... }(n) ~=~ \sum_{m=1}^n \frac{(-1)^m}{m^a} S_{-b, ... }(m).
\label{B4}
\end{eqnarray}

To demonstrate the nonsmooth behavior of such type of the nested sums,
we show the  sum $S_{-3, -1, 1, 1}(n)$ in Fig.~\ref{sa-bcds-a-bcd}.

Using the Eq. (\ref{A9.a}), we can express the function
$(-1)^m S_{-b, ... }(m)$ in the r.h.s. of (\ref{B4})
as a combination of the smooth function
$ \overline S_{-b, ... }^+(m)$ and some simpler functions and, thus, we have
\begin{eqnarray}
S_{-a, -b, ... }(n) &=& \sum_{m=1}^n \frac{1}{m^a}
\biggl[ \overline S^+_{-b, ... }(m) -
(1-(-1)^m)  Z_{-b, ... } \biggr]  \nonumber \\
 &=&  \sum_{m=1}^n \frac{1}{m^a}   \overline S^+_{-b, ... }(m)
+   Z_{-b, ... }  \biggl[S_{-a}(n)-S_{a}(n)\biggr].
\label{B5}
\end{eqnarray}
One can see that only one function $S_{-a}(n)$ is nonsmooth one in the
r.h.s. of (\ref{B5}).

Then, we have for the analytical continuation of $S_{-a, -b, ... }(n)$:
\begin{eqnarray}
\overline S^+_{-a, -b, ... }(n) ~=~
 \sum_{m=1}^n \frac{1}{m^a}   \overline S^+_{-b, ... }(m) +   Z_{-b, ... }
 \biggl[\overline S^+_{-a}(n)-S_{a}(n)\biggr].
\label{B6}
\end{eqnarray}

Taking the difference of Eqs. (\ref{B5}) and (\ref{B6}), we obtain
\begin{eqnarray}
\overline S^+_{-a, -b, ... }(n) - S_{-a, -b, ... }(n) &=&
  Z_{-b, ... }  \biggl[\overline S^+_{-a}(n)-S_{-a}(n)\biggr]
\nonumber  \\
 &=&  (1-(-1)^n)   Z_{-b, ... }  \biggl[Z_{-a}-S_{-a}(n)\biggr].
\label{B7}
\end{eqnarray}

From definitions (\ref{I0}) and (\ref{I0ad}) (by analogy
with Eqs. (\ref{Z1}) and  (\ref{Z4})) we have that
\begin{eqnarray} \z
Z_{-a,-b} ~=~ \zeta(-a,-b) +  \zeta(a+b), \nonumber \\
\z
Z_{-a,-b,c} ~=~ \zeta(-a,-b,c) +  \zeta(a+b,c) + \zeta(-a,-(b+c)) +
\zeta(a+b+c), \nonumber \\
\z
Z_{-a,-b,c,d} ~=~ \zeta(-a,-b,c,d) +  \zeta(a+b,c,d) +
\zeta(-a,-(b+c),d) +  \zeta(-a,-b,c+d)  \nonumber \\
\z \vspace{1cm}
+ \zeta(a+b+c,d) +  \zeta(a+b,c+d) + \zeta(-a,-(b+c+d)) +
\zeta(a+b+c+d).
\label{Z5}
\end{eqnarray}

The results for $S_{-3, -1, 1, 1}(n)$
are presented in Fig.~\ref{sa-bcds-a-bcd}, where the function
$\overline S^+_{-3, -1, 1, 1}(n)$ demonstrates its
smooth $n$ behavior.

  {\bf 3}. Now consider the sum $S_{a, -b, ... }(n)$, which
coincide in the case of two subscripts with one studied already in the
subsection  {\bf 1}.

To demonstrate the nonsmooth behavior of such type of the nested sums,
we show the  sum $S_{2, -1, 2, 1}(n)$ in Fig.~\ref{sa-bcds-a-bcd}.

By analogy with the previous subsection we can represent
the function $S_{a, -b, ... }(n)$ to the form
\begin{eqnarray}
S_{a, -b, ... }(n) ~=~ \sum_{m=1}^n \frac{1}{m^a} S_{-b, ... }(m)
~=~ \sum_{m=1}^n \frac{(-1)^m}{m^a} (-1)^m S_{-b, ... }(m).
\label{B8}
\end{eqnarray}

Using the Eq. (\ref{A9.a}) we can express  the function
$(-1)^m S_{-b, ... }(m)$ as a combination of the smooth function
$ \overline S_{-b, ... }^+(m)$ and some simpler functions
\begin{eqnarray}
S_{a, -b, ... }(n) &=& \sum_{m=1}^n \frac{(-1)^m}{m^a}
\biggl[  \overline S^+_{-b, ... }(m) -
(1-(-1)^m)  Z_{-b, ... } \biggr]  \nonumber \\
 &=&  \sum_{m=1}^n \frac{(-1)^m}{m^a} \overline S^+_{-b, ... }(m)
+   Z_{-b, ... }  \biggl[S_{a}(n)-S_{-a}(n)\biggr].
\label{B9}
\end{eqnarray}

One can see that the function $S_{-a}(n)$ is nonsmooth. Moreover, the
first term in the r.h.s. contains the smooth function
$ \overline S_{-b, ... }^+(m)$ and, thus, it can be continuated to odd
$n$ values by analogy
with the Eq. (\ref{A9.a}).

Then, we have the analytical continuation of $S_{a, -b, ... }(n)$ as
\begin{eqnarray}
\overline S^+_{a, -b, ... }(n) &=&
(-1)^n  \sum_{m=1}^n \frac{(-1)^m}{m^a} \overline S^+_{-b, ... }(m) +
(1-(-1)^n)  \sum_{m=1}^{\infty} \frac{(-1)^m}{m^a}
\overline S^+_{-b, ... }(m)   \nonumber \\
&+&  Z_{-b, ... }
 \biggl[S_{a}(n) - \overline S^+_{-a}(n)\biggr].
\label{B10}
\end{eqnarray}

Using Eqs. (\ref{A7}) and (\ref{A9.a}) to represent the functions
$\overline S^+_{-a}(n)$ and $\overline S^+_{-b, ... }(m)$
as combinations of $S_{-a}(n)$,  $Z_{-a}$, $S_{-b, ... }(m)$
and   $Z_{-b, ... }$, we have the final result
\begin{eqnarray}
\overline S^+_{a, -b, ... }(n) ~=~
(-1)^n S_{a, -b, ... }(n) +
 (1-(-1)^n)  \biggl[  Z_{a,-b, ... } -  Z_{-b, ... }
\Bigl( Z_{a}-S_{a}(n) \Bigr) \biggr].
\label{B11}
\end{eqnarray}

From definitions (\ref{I0}) and (\ref{I0ad}) we have that
\begin{eqnarray} \z
Z_{a,-b} ~=~ \zeta(a,-b) +  \zeta(-(a+b)), \nonumber \\
\z
Z_{a,-b,c} ~=~ \zeta(a,-b,c) +  \zeta(-(a+b),c) + \zeta(a,-(b+c)) +
\zeta(-(a+b+c)), \nonumber \\
\z
Z_{a,-b,c,d} ~=~ \zeta(a,-b,c,d) +  \zeta(-(a+b),c,d) +
\zeta(a,-(b+c),d)+ \zeta(-(a+b+c),d)  \nonumber \\
\z \vspace{1cm}
 +  \zeta(a,-b,c+d) +  \zeta(-(a+b),c+d) + \zeta(a,-(b+c+d)) +
\zeta(-(a+b+c+d)).
\label{Z6}
\end{eqnarray}
One can see that the results coincide with the Eq.  (\ref{B3}) in the
case of two subscripts.

The results for $S_{2, -1, 2, 1}(n)$
are presented in Fig.~\ref{sa-bcds-a-bcd}, where the function
$\overline S^+_{2, -1, 2, 1}(n)$ demonstrates its
smooth $n$ behavior.\\

  {\bf 4}. Consider the sum $S_{a, -b, -c, ... }(n)$. By analogy with the
subsection  {\bf 2} we have
\begin{eqnarray}
S_{a, -b, -c, ... }(n) ~=~ \sum_{m=1}^n \frac{1}{m^a} S_{-b,-c, ... }(m).
\label{B12}
\end{eqnarray}

\begin{figure}[t]
\begin{center}
\epsfig{figure=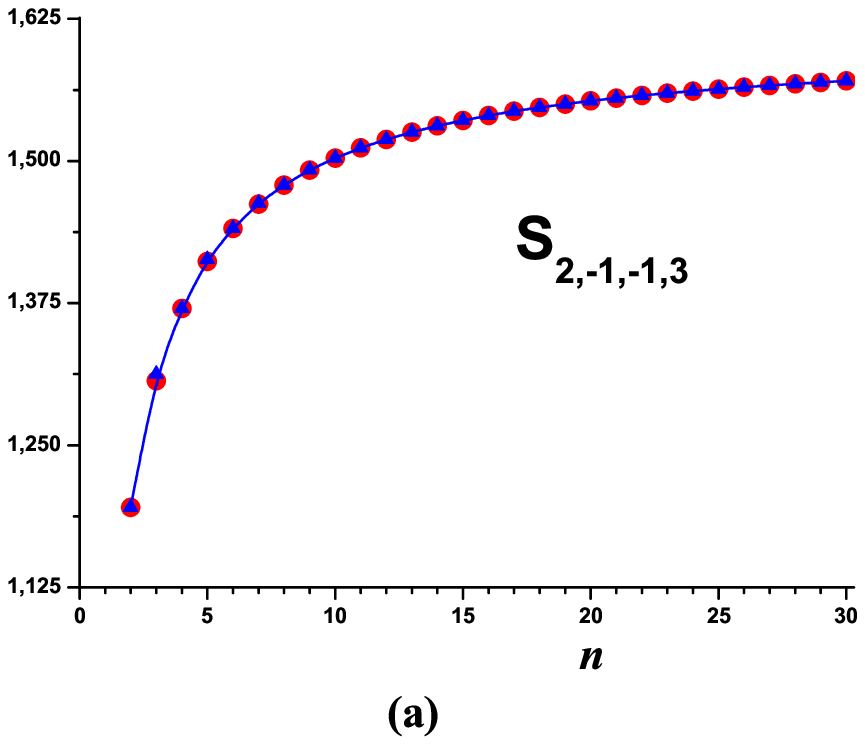,width=8cm}\epsfig{figure=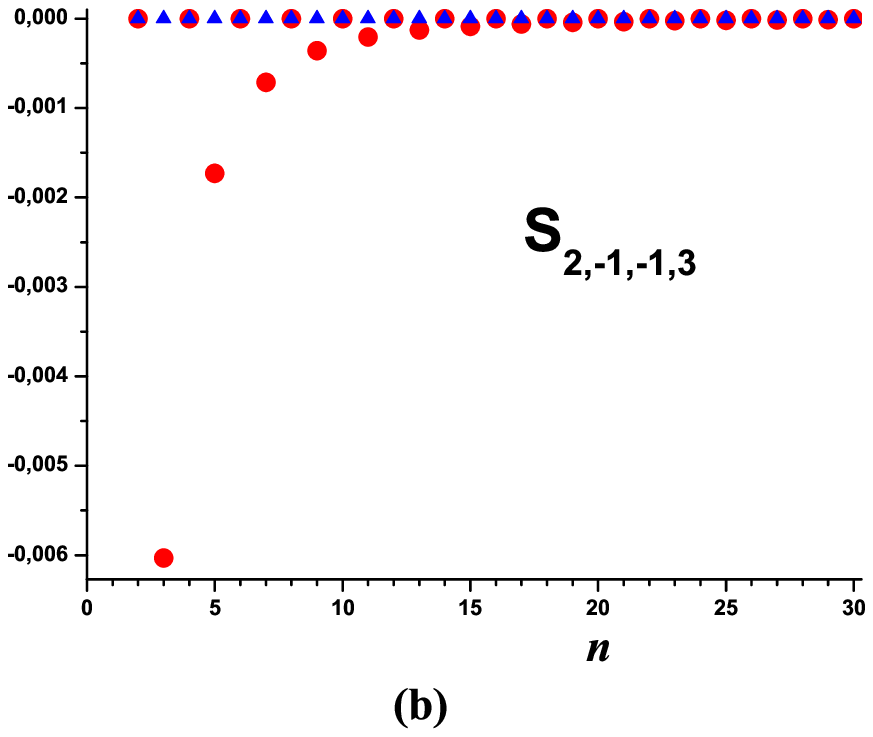,width=8cm}
\end{center}
\caption{The circles are represented the sum $S_{2,-1,-1,3}(n)$
and
the difference $\overline S_{2,-1,-1,3}^+(n)-S_{2,-1,-1,3}(n)$
(in the parts (a) and (b), respectively).
}
\label{sa-b-cd}
\end{figure}

Using the Eq. (\ref{B7}), we can express the function
$S_{-b,-c, ... }(m)$ as a combination of the smooth function
$ \overline S_{-b,-c, ... }^+(m)$ and some simpler functions:
\begin{eqnarray}
S_{a, -b, -c, ... }(n) &=& \sum_{m=1}^n \frac{1}{m^a}
\biggl[ \overline S^+_{-b,-c, ... }(m) -
(1-(-1)^m)  Z_{-c, ... } \Bigl[ Z_{-b}-S_{-b}(m)\Bigr]
 \biggr]  \nonumber \\
&& \hspace{-4cm} ~=~
 \sum_{m=1}^n \frac{1}{m^a}   \overline S^+_{-b,-c, ... }(m)
+   Z_{-c, ... }  \biggl[  Z_{-b} \Bigl(S_{-a}(n)-S_{a}(n)\Bigr)
-S_{-a,-b}(n)+S_{a,-b}(n) \biggr].
\label{B13}
\end{eqnarray}
One can see that only the functions $S_{-a}(n)$, $S_{-a,-b}(n)$ and
$S_{a,-b}(n)$ are nonsmooth one.

Then, we have for the analytical continuation of $S_{a, -b, -c, ... }(n)$:
\begin{eqnarray}
\overline S^+_{a, -b, -c, ... }(n) &=&
 \sum_{m=1}^n \frac{1}{m^a}   \overline S^+_{-b, -c ... }(m) +
Z_{-c, ... }  \biggl[  Z_{-b} \Bigl(\overline S_{-a}^+(n)-S_{a}(n)\Bigr)
 \nonumber \\
&-& \overline S^+_{-a,-b}(n)+\overline S^+_{a,-b}(n) \biggr].
\label{B14}
\end{eqnarray}
Taking the difference of Eqs. (\ref{B13}) and (\ref{B14}), we obtain
\begin{eqnarray}
\overline S^+_{a, -b,-c, ... }(n) &-& S_{a, -b,-c, ... }(n) ~=~
Z_{-c, ... }  \biggl[  Z_{-b} \Bigl(\overline S_{-a}^+(n)-S_{-a}(n)\bigr)
 \nonumber \\
&-&\Bigl( \overline S^+_{-a,-b}(n)- S_{-a,-b}(n)\Bigr) + \Bigl(
\overline S^+_{a,-b}(n)- S_{a,-b}(n) \Bigr) \biggr]
\nonumber  \\
 &=&  (1-(-1)^n)   Z_{-c, ... }  \biggl[Z_{a,-b}-S_{a,-b}(n) -Z_{-b}
\Bigr( Z_{a}-S_{a}(n)\Bigr)\biggr].
\label{B15}
\end{eqnarray}

From definitions (\ref{I0}) and (\ref{I0ad}) we have that
\begin{eqnarray} \z
Z_{a,-b,-c} ~=~ \zeta(a,-b,-c) +  \zeta(-(a+b),-c) + \zeta(a,b+c) +
\zeta(a+b+c), \nonumber \\
\z
Z_{a,-b,-c,d} ~=~ \zeta(a,-b,-c,d) +  \zeta(-(a+b),-c,d) +
\zeta(a,b+c,d) +  \zeta(a,-b,-(c+d))  \nonumber \\
\z \vspace{1cm}
+ \zeta(a+b+c,d) +  \zeta(-(a+b),-(c+d)) + \zeta(a,b+c+d) +
\zeta(a+b+c+d).
\label{Z7}
\end{eqnarray}

As an example, we show the sum $S_{2, -1, -1, 3}(n)$
in Fig.~\ref{sa-b-cd}(a). For the nested sums, where the first index $a$
is positive, the difference between two function, which are generated
at even and odd $n$ values, are not so strong (see also, for example,
the nested sums $S_{-3, -1, 1, 1}(n)$ and $S_{2, -1, -1, 3}(n)$ in
Fig.~\ref{sa-bcds-a-bcd}).

To demonstrate the effect of the analytic continuation, we show in
Fig.~\ref{sa-b-cd}(b) (by circles) the difference between
$S_{2, -1, -1, 3}(n)$ and the function, which is approximated from even
$n$ values to integer ones.
We see the effect of the difference at the odd $n$ values (essentially at
$n=3,5,7$). After the analytic
continuation, the difference become to be zero, that it is shown by
triangles in Fig.~\ref{sa-b-cd}(b).\\

\begin{figure}[t]
\begin{center}
\epsfig{figure=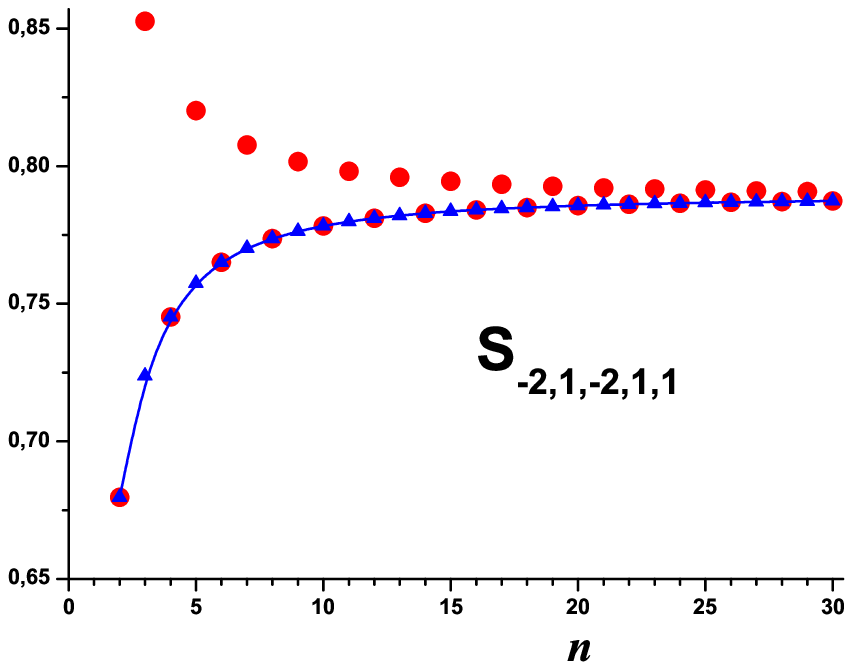,width=8cm}\epsfig{figure=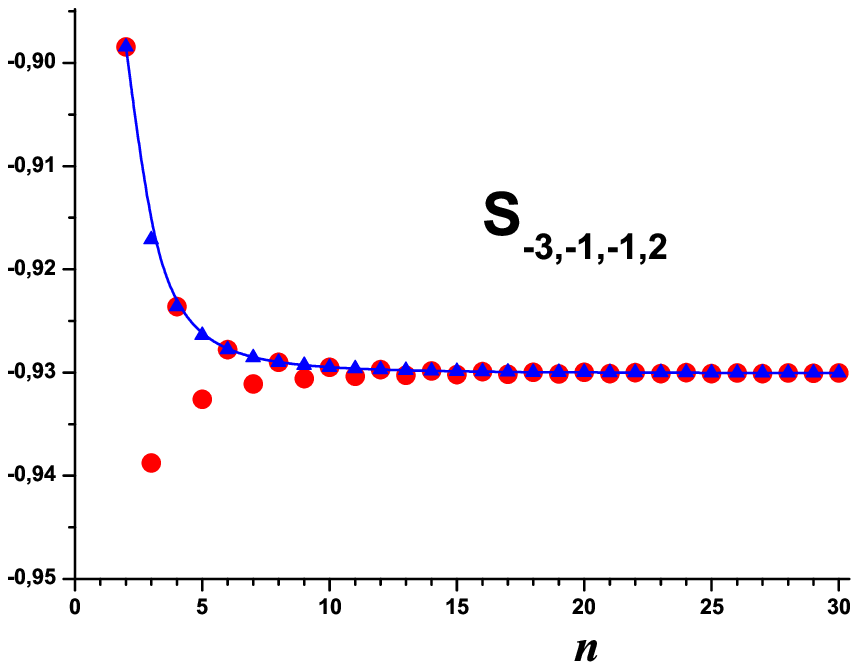,width=8cm}
\end{center}
\caption{As in Fig.~\ref{s-a} but for the sum $S_{-2, 1, -2, 1, 1}(n)$
and $S_{-3, -1, -1, 2}(n)$.}
\label{s-ab-cde}
\end{figure}

  {\bf 5}. Consider the sum $S_{-a, b, -c, ... }(n)$.
To demonstrate the nonsmooth behavior of such type of the nested sums,
we show the  sum $S_{-2, 1, -2, 1, 1}(n)$ in Fig. ~\ref{s-ab-cde}.

By analogy with the subsections  {\bf 2} and {\bf 4} we have
\begin{eqnarray}
S_{-a, b, -c, ... }(n) ~=~ \sum_{m=1}^n \frac{(-1)^m}{m^a} S_{b,-c, ... }(m).
\label{B16}
\end{eqnarray}

Using the Eq. (\ref{B11}), we can express the function
$(-1)^m S_{b,-c, ... }(m)$ as a combination of the smooth function
$ \overline S_{b,-c, ... }^+(m)$ and some simpler functions:
\begin{eqnarray}
S_{-a, b, -c, ... }(n) &=& \sum_{m=1}^n \frac{1}{m^a}
\biggl[ \overline S^+_{b,-c, ... }(m) -
(1-(-1)^m) \biggl(  Z_{b,-c, ... } - Z_{-c, ... } \Bigl[ Z_{b}-S_{b}(m)\Bigr]
\biggr)  \biggr]  \nonumber \\
&=&
 \sum_{m=1}^n \frac{1}{m^a}   \overline S^+_{b,-c, ... }(m)
+  \biggl(  Z_{b,-c, ... } -   Z_{b} Z_{-c, ... } \biggr)
\Bigl[S_{-a}(n)-S_{a}(n)\Bigr] \nonumber \\
&+& Z_{-c, ... }
\biggr[S_{-a,b}(n)-S_{a,b}(n) \biggr].
\label{B17}
\end{eqnarray}
One can see that only the functions $S_{-a}(n)$ and
$S_{-a,b}(n)$ are nonsmooth one.

Then, we have for the analytical continuation of $S_{-a, b, -c, ... }(n)$:
\begin{eqnarray}
\overline S^+_{-a, b, -c, ... }(n) &=&
 \sum_{m=1}^n \frac{1}{m^a}   \overline S^+_{b, -c ... }(m) +
 \biggl(  Z_{b,-c, ... } -  Z_{b}  Z_{-c, ... } \biggr)
\Bigl[\overline S^+_{-a}(n)-S_{a}(n)\Bigr]  \nonumber \\
& +& Z_{-c, ... }
\biggr[\overline S^+_{-a,b}(n)-S_{a,b}(n) \biggr].
\label{B18}
\end{eqnarray}
Taking the difference of Eqs. (\ref{B17}) and (\ref{B18}), we obtain
\begin{eqnarray}
\overline S^+_{-a, b,-c, ... }(n) &-& S_{-a, b,-c, ... }(n) ~=~
\biggl(  Z_{b,-c, ... } -   Z_{b} Z_{-c, ... } \biggr)
\Bigl[\overline S^+_{-a}(n)-S_{-a}(n)\Bigr]  \nonumber \\
&+& Z_{-c, ... }
\biggr[\overline S^+_{-a,b}(n)-S_{-a,b}(n) \biggr]
\nonumber  \\
&& \hspace{-4cm} ~=~
(1-(-1)^n)
\Biggl( \biggl[ Z_{b,-c, ... } -   Z_{b} Z_{-c, ... }  \biggr]
\Bigl[Z_{-a}-S_{-a}(n)\Bigr]
+ Z_{-c, ... }
\biggr[Z_{-a,b}-S_{-a,b}(n) \biggr] \Biggr).
\label{B19}
\end{eqnarray}
From definitions (\ref{I0}) and (\ref{I0ad}) we have that
\begin{eqnarray} \z\ \
Z_{-a,b,-c} = \zeta(-a,b,-c) +  \zeta(-(a+b),-c) + \zeta(-a,-(b+c))
+  \zeta(a+b+c), \nonumber \\
\z\ \
Z_{-a,b,-c,d} = \zeta(-a,b,-c,d) +  \zeta(-(a+b),-c,d) +
\zeta(-a,-(b+c),d) +  \zeta(-a,b,-(c+d))  \nonumber \\
\z\ \  \vspace{1cm}
+ \zeta(a+b+c,d) +  \zeta(-(a+b),-(c+d)) + \zeta(-a,-(b+c+d)) +
\zeta(a+b+c+d).
\label{Z8}
\end{eqnarray}

The results for $S_{-2, 1, -2, 1, 1}(n)$
are presented in Fig.~\ref{s-ab-cde}, where the function
$\overline S^+_{-2, 1, -2, 1, 1}(n)$ demonstrates its
smooth $n$ behavior.\\

  {\bf 6}. Consider the sum $S_{a, b, -c, ... }(n)$

\begin{figure}[t]
\begin{center}
\epsfig{figure=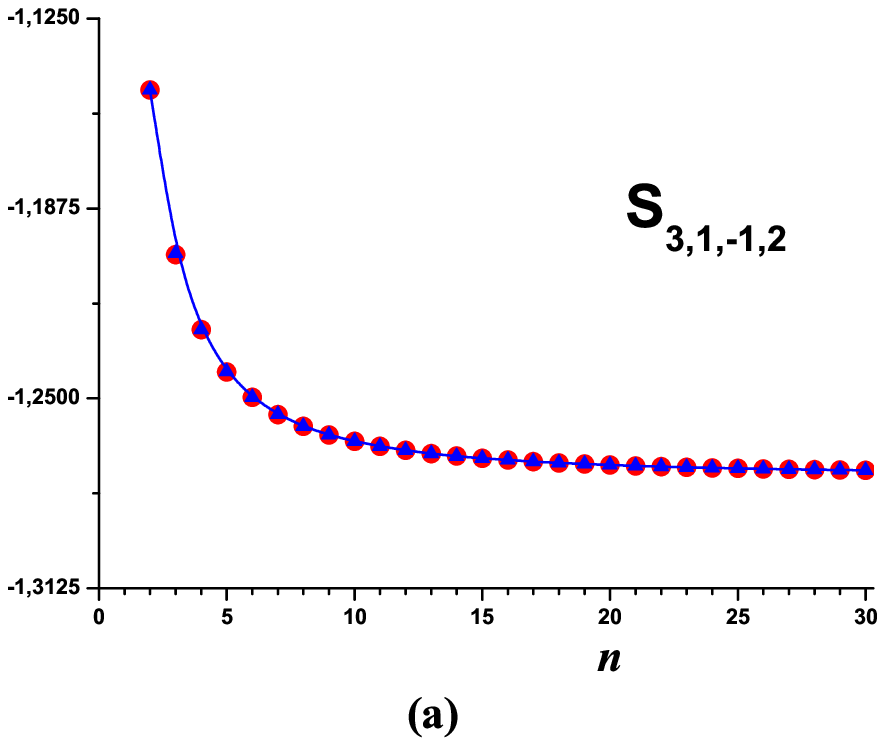,width=8cm}\epsfig{figure=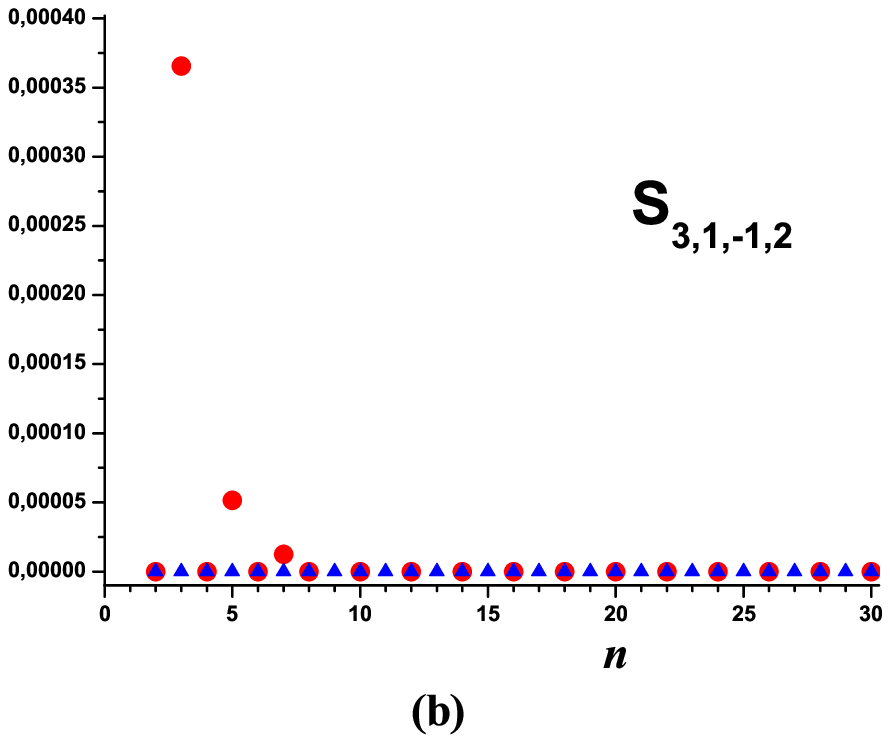,width=8cm}
\end{center}
\caption{As in Fig.~\ref{sa-b-cd} but for the sum $S_{3, 1, -1, 2}(n)$.}
\label{sab-cd}
\end{figure}

By analogy with the subsections  {\bf 3} and {\bf 5} we have
\begin{eqnarray}
S_{a, b, -c, ... }(n) ~=~ \sum_{m=1}^n \frac{1}{m^a} S_{b,-c, ... }(m)
~=~ \sum_{m=1}^n \frac{(-1)^m}{m^a} (-1)^m S_{b,-c, ... }(m).
\label{B20}
\end{eqnarray}

Using the Eq. (\ref{B11}), we can express the function
$(-1)^m S_{b,-c, ... }(m)$ as a combination of the smooth function
$ \overline S_{b,-c, ... }^+(m)$ and some simpler functions:
\begin{eqnarray}
S_{a, b, -c, ... }(n) &=& \sum_{m=1}^n \frac{(-1)^m}{m^a}
\biggl[ \overline S^+_{b,-c, ... }(m) -
(1-(-1)^m) \biggl(  Z_{b,-c, ... } - Z_{-c, ... } \Bigl[ Z_{b}-S_{b}(m)\Bigr]
\biggr)  \biggr]  \nonumber \\
&=&
 \sum_{m=1}^n \frac{(-1)^m}{m^a}   \overline S^+_{b,-c, ... }(m)
+  \biggl(  Z_{b,-c, ... } -  Z_{b} Z_{-c, ... }  \biggr)
\Bigl[S_{a}(n)-S_{-a}(n)\Bigr] \nonumber \\
&+& Z_{-c, ... }
\biggr[S_{a,b}(n)-S_{-a,b}(n) \biggr].
\label{B21}
\end{eqnarray}

One can see that the functions $S_{-a}(n)$ and
$S_{-a,b}(n)$ are nonsmooth one. Moreover, the
first term in the r.h.s. contains the smooth function
$ \overline S_{b,-c, ... }^+(m)$ and, thus, can be continuated by analogy
with the Eq. (\ref{B11}).

Then, we have for the analytical continuation of $S_{a, b, -c, ... }(n)$:
\begin{eqnarray}
\overline S^+_{a, b, -c, ... }(n) &=&  (-1)^n
 \sum_{m=1}^n \frac{(-1)^m}{m^a}   \overline S^+_{b, -c ... }(m) +
(1-(-1)^n)
\sum_{m=1}^{\infty} \frac{(-1)^m}{m^a}   \overline S^+_{b, -c ... }(m)
 \nonumber \\
&& \hspace{-2cm} -
 \biggl(  Z_{b,-c, ... } -  Z_{b} Z_{-c, ... }  \biggr)
\Bigl[\overline S^+_{-a}(n)-S_{a}(n)\Bigr]
- Z_{-c, ... }
\biggr[\overline S^+_{-a,b}(n)-S_{a,b}(n) \biggr].
\label{B22}
\end{eqnarray}

Using Eqs. (\ref{A7}), (\ref{A9.a}) and (\ref{B11}) to represent the functions
$\overline S^+_{-a}(n)$, $\overline S^+_{-a,b, ... }(n)$ and
$\overline S^+_{b,-c, ... }(m)$
as combinations of $S_{-a}(n)$,  $Z_{-a}$,
$S_{-a,b, ... }(n)$,   $Z_{-a,b, ... }$,
$S_{b,-c, ... }(m)$ and   $Z_{b,-c, ... }$.

After some algebra we have the final result
\begin{eqnarray}
\overline S^+_{a, b,-c, ... }(n) &=& (-1)^n S_{a, b,-c, ... }(n)
~+~
(1-(-1)^n)
\Biggl( Z_{a, b,-c, ... } \nonumber  \\
&-&
\biggl[ Z_{b,-c, ... } -  Z_{b} Z_{-c, ... }  \biggr]
\Bigl[Z_{a}-S_{a}(n)\Bigr]
- Z_{-c, ... }
\biggr[Z_{a,b}-S_{a,b}(n) \biggr] \Biggr).
\label{B23}
\end{eqnarray}
From definitions (\ref{I0}) and (\ref{I0ad}) we have that
\begin{eqnarray} \z\hspace*{-6mm}
Z_{a,b,-c} ~=~ \zeta(a,b,-c) +  \zeta(a+b,-c) + \zeta(a,-(b+c))
+  \zeta(-(a+b+c)), \nonumber \\
\z\hspace*{-6mm}
Z_{a,b,-c,d} ~=~ \zeta(a,b,-c,d) +  \zeta(a+b,-c,d) +
\zeta(a,-(b+c),d) + \zeta(-(a+b+c),d)  \nonumber \\
\z\hspace*{-5mm} \vspace{1cm}
+  \zeta(a,b,-(c+d)) +  \zeta(a+b,-(c+d)) + \zeta(a,-(b+c+d)) +
\zeta(-(a+b+c+d)).
\label{Z9}
\end{eqnarray}

As an example, we show the sum $S_{3, 1, -1, 2}(n)$
in Fig.~\ref{sab-cd}(a). By analogy with the subsection {\bf 4},
to demonstrate the effect of the analytic continuation, we show in
Fig.~\ref{sab-cd}(b) (by circles) the difference between
$S_{3, 1, -1, 2}(n)$ and the function, which is approximated from even
$n$ values to integer ones.
We see the effect of the difference at the odd $n$ values (essentially at
$n=3,5,7$). After the analytic
continuation, the difference become to be zero, that it is shown by
triangles in Fig.~\ref{sab-cd}(b).\\

  {\bf 7}. As a last one we consider the sum $S_{-a, -b, -c, ... }(n)$.
To demonstrate the nonsmooth behavior of such type of the nested sums,
we show the  sum $S_{-3, 1, -1, 2}(n)$ in Fig. ~\ref{s-ab-cde}.

By analogy with the subsections  {\bf 2} and {\bf 4} we have
\begin{eqnarray}
S_{-a, -b, -c, ... }(n) ~=~ \sum_{m=1}^n \frac{(-1)^m}{m^a} S_{-b,-c, ... }(m).
\label{B24}
\end{eqnarray}

Using the Eq. (\ref{B7}), we can express the function
$S_{-b,-c, ... }(m)$ as a combination of the smooth function
$ \overline S_{-b,-c, ... }^+(m)$ and some simpler functions:
\begin{eqnarray}
S_{-a, -b, -c, ... }(n) &=& \sum_{m=1}^n \frac{(-1)^m}{m^a}
\biggl[ \overline S^+_{-b,-c, ... }(m) -
(1-(-1)^m)  Z_{-c, ... } \Bigl[ Z_{-b}-S_{-b}(m)\Bigr]
 \biggr]  \nonumber \\
&& \hspace{-45mm} ~=~
 \sum_{m=1}^n \frac{(-1)^m}{m^a}   \overline S^+_{-b,-c, ... }(m)
+   Z_{-c, ... }  \biggl[  Z_{-b} \Bigl(S_{a}(n)-S_{-a}(n)\Bigr)
+S_{-a,-b}(n)-S_{a,-b}(n) \biggr].
\label{B25}
\end{eqnarray}

One can see that the functions $S_{-a}(n)$, $S_{-a,-b}(n)$ and
$S_{a,-b}(n)$ are nonsmooth. Moreover, the
first term in the r.h.s. contains the smooth function
$ \overline S_{-b,-c, ... }^+(m)$ and, thus, can be continuated by analogy
with the Eq. (\ref{B7}).

Then, we have for the analytical continuation of $S_{a, -b, -c, ... }(n)$:
\begin{eqnarray}
\overline S^+_{-a, -b, -c, ... }(n) &=&
(-1)^n  \sum_{m=1}^n \frac{(-1)^m}{m^a} \overline S^+_{-b,-c, ... }(m) +
(1-(-1)^n)  \sum_{m=1}^{\infty} \frac{(-1)^m}{m^a}
\overline S^+_{-b,-c, ... }(m)  \nonumber \\
&-&
Z_{-c, ... }  \biggl[  Z_{-b} \Bigl(\overline S_{-a}^+(n)-S_{a}(n)\Bigr)
- \overline S^+_{-a,-b}(n)+\overline S^+_{a,-b}(n) \biggr].
\label{B26}
\end{eqnarray}

Using Eqs. (\ref{A7}), (\ref{A9.a}) and (\ref{B7}) to represent the functions
$\overline S^+_{-a}(n)$, $\overline S^+_{a,-b}(n)$ and $\overline S^+_{-b, -c, ... }(m)$
as combinations of $S_{-a}(n)$,  $Z_{-a}$, $S_{-b, ... }(m)$
and   $Z_{-b, ... }$, we have the final result
\begin{eqnarray}
\z\overline S^+_{-a, -b, -c, ... }(n) \ =\
(-1)^n S_{-a, -b, -c, ... }(n)  \nonumber \\
\z\ -\  (1-(-1)^n)   Z_{-c, ... }  \biggl[Z_{-a,-b}-S_{-a,-b}(n) -Z_{-b}
\Bigr( Z_{-a}-S_{-a}(n)\Bigr)\biggr].
\label{B27}
\end{eqnarray}
From definitions (\ref{I0}) and (\ref{I0ad}) we have that
\begin{eqnarray} \z
Z_{-a,-b,-c} ~=~ \zeta(-a,-b,-c) +  \zeta(a+b,-c) + \zeta(a,b+c)
+  \zeta(-(a+b+c)), \nonumber \\
\z
Z_{-a,-b,-c,d} ~=~ \zeta(-a,-b,-c,d) +  \zeta(a+b,-c,d) +
\zeta(-a,b+c,d) +  \zeta(-a,-b,-(c+d))  \nonumber \\
\z \vspace{1cm}
\,+\, \zeta(-(a+b+c),d)\! +\!  \zeta(a+b,-(c+d))\! +\! \zeta(-a,-b+c+d)\! +\!
\zeta(-(a+b+c+d)).
\label{Z9a}
\end{eqnarray}

The results for $S_{-3, 1, -1, 2}(n)$
are presented in Fig.~\ref{s-ab-cde}, where the function
$\overline S^+_{-3, 1, -1, 2}(n)$ demonstrates its
smooth $n$ behavior.

\section{ Appendix B}
\def\theequation{B\arabic{equation}}
\label{App:B}
\setcounter{equation}0

Here we derive the analytic continuation of the nested sums
$$S_{a, -b, c, ... }(n), ~ S_{-a, -b, c, ... }(n),~
S_{a, -b, -c, ... }(n),~ S_{-a, b, -c, ... }(n),~ S_{a, b, -c, ... }(n),~
S_{-a, -b, -c, ... }(n)$$
from the even $n$ values to the real and/or complex $n$ values
(the final results are presented in Section 4). As it was in Appendix A
we will use here the definition (\ref{B0}).\\

  {\bf 1}. Consider firstly the sum $S_{-a, -b, ... }(n)$.
Using Eq.~(\ref{B6}) we rewrite the first term at the r.h.s. as follows
\begin{eqnarray}
\sum_{m=1}^n \frac{1}{m^a} \overline S_{-b, ... }^{+}(m) ~=~
\biggl[\sum_{m=1}^{\infty} -\sum_{m=n+1}^{\infty}\biggr]
 \frac{1}{m^a} \overline S_{-b, ... }^{+}(m).
\label{C1}
\end{eqnarray}

The first term at the r.h.s. of (\ref{C1}) is equal to
\begin{eqnarray}
\sum_{m=1}^{\infty} \frac{1}{m^a} \overline S_{-b, ... }^{+}(m) ~=~
Z_{-a,-b, ... } + Z_{-b, ... } \Bigl[ Z_{a}- Z_{-a}\Bigr].
\label{C2}
\end{eqnarray}

The second term can be defined as
\begin{eqnarray}
\sum_{m=n+1}^{\infty} \frac{1}{m^a} \overline S_{-b, ... }^{+}(m) ~=~
\sum_{l=0}^{\infty} \frac{1}{(l+n+1)^a} \overline S_{-b,
  ... }^{+}(l+n+1) ~\equiv~
\Psi_{a,-b, ... }(n+1),
\label{C3}
\end{eqnarray}
where the function $ \Psi_{a,-b, ... }(n+1)$ is well defined for
the real and/or complex $n$ values.

Taking together Eqs. (\ref{A8}), (\ref{B6}) and (\ref{C1})-(\ref{C3}),
we obtain the following result
\begin{eqnarray}
\overline S_{-a,-b, ... }^{+}(n) ~=~ Z_{-a,-b, ... } -
\Psi_{a,-b, ... }(n+1)
+ Z_{-b, ... } \Bigl[ \Psi_{a}(n+1)- \Psi_{-a}(n+1)\Bigr],
\label{C4}
\end{eqnarray}
i.e. now the function $\overline S_{-a,-b, ... }^{+}(n)$
contains objects well defined for
the real and/or complex $n$ values.\\

  {\bf 2}. Consider now the sum $S_{a, -b, ... }(n)$.
By analogy with the previous subsection,
we rewrite the first term at the r.h.s. of Eq.~(\ref{B10})
as follows
\begin{eqnarray}
\z
\sum_{m=1}^n \frac{(-1)^m}{m^a} \overline S_{-b, ... }^{+}(m) ~=~
\sum_{m=1}^{\infty}
 \frac{(-1)^m}{m^a} \overline S_{-b, ... }^{+}(m) - \sum_{l=0}^{\infty}
\frac{(-1)^{n+l+1}}{(l+n+1)^a} \overline S_{-b, ... }^{+}(l+n+1) \nonumber \\
\z =~
Z_{a,-b, ... } + Z_{-b, ... } \Bigl[ Z_{-a}- Z_{a}\Bigr]
- (-1)^n \, \Psi_{-a,-b, ... }(n+1).
\label{C5}
\end{eqnarray}

Taking together Eqs. (\ref{A8}), (\ref{B10}) and (\ref{C5}),
we obtain the following result
\begin{eqnarray}
\overline S_{a,-b, ... }^{+}(n) ~=~ Z_{a,-b, ... } -
\Psi_{-a,-b, ... }(n+1) +
 Z_{-b, ... } \Bigl[ \Psi_{-a}(n+1)- \Psi_{a}(n+1)\Bigr],
\label{C6}
\end{eqnarray}
i.e. now the function $\overline S_{a,-b, ... }^{+}(n)$ is well defined for
the real and/or complex $n$ values.\\

  {\bf 3}. Consider the sums $S_{a, -b, -c, ... }(n)$,
$S_{-a, b, -c, ... }(n)$, $S_{a, b, -c, ... }(n)$ and $S_{-a, -b, -c, ... }(n)$.
By analogy with the previous subsections,
the first term at the r.h.s. of Eqs.~(\ref{B14}), (\ref{B18}), (\ref{B22})
and (\ref{B26}) can be represented in the following form, respectively,
\begin{eqnarray}
\z \sum_{m=1}^n \frac{1}{m^a} \overline S_{-b,-c, ... }^{+}(m) ~=~
\sum_{m=1}^{\infty}
 \frac{1}{m^a} \overline S_{-b,-c, ... }^{+}(m) - \sum_{l=0}^{\infty}
\frac{1}{(l+n+1)^a} \overline S_{-b,-c ... }^{+}(l+n+1) \nonumber \\
\z\qquad =~ Z_{a,-b,-c, ... } + Z_{-c, ... } \biggr[ Z_{-b}
 \Bigl( Z_{a}- Z_{-a}\Bigr) -Z_{a,-b}+ Z_{-a,-b}\biggr] -
\Psi_{a,-b,-c ... }(n+1),
\label{C7}\\
\z \nonumber \\
\z \sum_{m=1}^n \frac{1}{m^a} \overline S_{b,-c, ... }^{+}(m)\ =\
\sum_{m=1}^{\infty}
 \frac{1}{m^a} \overline S_{b,-c, ... }^{+}(m) - \sum_{l=0}^{\infty}
\frac{1}{(l+n+1)^a} \overline S_{b,-c ... }^{+}(l+n+1) \nonumber \\
\z\qquad =\ Z_{-a,b,-c, ... } + \Bigl[Z_{b,-c, ... } -Z_{b}Z_{-c, ... }\Bigr]
 \Bigl( Z_{a}- Z_{-a}\Bigr) + Z_{-c, ... } \Bigl(
Z_{a,b}- Z_{-a,b}\Bigr)
 \nonumber \\
\z\qquad -\
\Psi_{a,b,-c ... }(n+1),
\label{C8}\\
\z \nonumber \\
\z \sum_{m=1}^n \frac{(-1)^m}{m^a} \overline S_{b,-c ... }^{+}(m) ~=~
\sum_{m=1}^{\infty}
 \frac{(-1)^m}{m^a} \overline S_{b,-c, ... }^{+}(m) - \sum_{l=0}^{\infty}
\frac{(-1)^{n+l+1}}{(l+n+1)^a} \overline S_{b,-c, ... }^{+}(l+n+1) \nonumber \\
\z\qquad =\ Z_{a,b,-c, ... } + \Bigl[Z_{b,-c, ... } -Z_{b}Z_{-c, ... }\Bigr]
 \Bigl( Z_{-a}- Z_{a}\Bigr) + Z_{-c, ... } \Bigl(
Z_{-a,b}- Z_{a,b}\Bigr) \nonumber \\
\z\qquad -\ \Psi_{-a,b,-c ... }(n+1).
\label{C7a}\\
\z \nonumber \\
\z \sum_{m=1}^n \frac{(-1)^m}{m^a} \overline S_{-b,-c, ... }^{+}(m) ~=~
\sum_{m=1}^{\infty}
 \frac{(-1)^m}{m^a} \overline S_{-b,-c, ... }^{+}(m) - \sum_{l=0}^{\infty}
\frac{(-1)^{l+n+1}}{(l+n+1)^a} \overline S_{-b,-c ... }^{+}(l+n+1) \nonumber \\
\z\qquad =~ Z_{-a,-b,-c, ... } + Z_{-c, ... } \biggr[ Z_{-b}
 \Bigl( Z_{-a}- Z_{a}\Bigr) +Z_{a,-b}- Z_{-a,-b}\biggr] -
\Psi_{-a,-b,-c ... }(n+1),
\label{C9}
\end{eqnarray}

Taking together Eqs. (\ref{A8}), (\ref{B14}), (\ref{B18}), (\ref{B22}), (\ref{B26})
and (\ref{C7})-(\ref{C9}),
we obtain the following results
\begin{eqnarray}
\z\ \ \overline S_{a,-b,-c, ... }^{+}(n) = Z_{a,-b,-c, ... } -
\Psi_{a,-b,-c, ... }(n+1)
~+~Z_{-c, ... } \biggl[ \Psi_{a,-b}(n+1)- \Psi_{-a,-b}(n+1) \nonumber \\
\z\qquad- Z_{-b} \Bigl( \Psi_{a}(n+1)- \Psi_{-a}(n+1)
\Bigr) \biggr],
\label{C10}\\
\z\ \ \overline S_{-a,b,-c, ... }^{+}(n) = Z_{-a,b,-c, ... } -
\Psi_{a,b,-c, ... }(n+1)
~+~ \Bigl[Z_{b,-c, ... } -Z_{b}Z_{-c, ... }\Bigr] \nonumber \\
\z\qquad\times\Bigl( \Psi_{a}(n+1)- \Psi_{-a}(n+1)\Bigr)
~+~Z_{-c, ... } \Bigl( \Psi_{a,b}(n+1)- \Psi_{-a,b}(n+1)\Bigr),
\label{C11}\\
\z\ \ \overline S_{a,b,-c, ... }^{+}(n) = Z_{a,b,-c, ... } -
\Psi_{-a,b,-c, ... }(n+1)
~+~ \Bigl[Z_{b,-c, ... } -Z_{b}Z_{-c, ... }\Bigr]\nonumber \\
\z\qquad\times \Bigl( \Psi_{-a}(n+1)-
\Psi_{a}(n+1)\Bigr)
~+~ Z_{-c, ... } \Bigl(\Psi_{-a,b}(n+1)- \Psi_{a,b}(n+1)\Bigr),
\label{C10a}\\
\z\ \ \overline S_{-a,-b,-c, ... }^{+}(n) = Z_{-a,-b,-c, ... } -
\Psi_{-a,-b,-c, ... }(n+1)
~+~Z_{-c, ... } \biggl[ \Psi_{-a,-b}(n+1)- \Psi_{a,-b}(n+1) \nonumber \\
\z\qquad+ Z_{-b} \Bigl( \Psi_{a}(n+1)- \Psi_{-a}(n+1)
\Bigr) \biggr],
\label{C10b}
\end{eqnarray}
i.e. now the functions
$\overline S_{a, -b, -c, ... }^{+}(n)$,
$\overline S_{-a, b, -c, ... }^{+}(n)$,
$\overline S_{a, b, -c, ... }^{+}(n)$ and
$\overline S_{-a, -b, -c, ... }^{+}(n)$
are well defined for
the real and/or complex $n$ values.

\end{document}